\documentclass[12pt,letterpaper]{article}
\usepackage[utf8]{inputenc}
\usepackage{amsmath}
\usepackage{mathptmx, courier,textcomp}
\usepackage{amsfonts}
\usepackage{amssymb}
\usepackage{graphicx}
\usepackage{cite}
\usepackage{ulem}
\usepackage{caption}
\usepackage{epstopdf}
\usepackage[top=1in,bottom=1in,left=1in,right=1in]{geometry} 
\usepackage{float}
\usepackage{hyperref}
\usepackage{booktabs}
\usepackage[export]{adjustbox}
\usepackage{float}
\usepackage{multirow}
\usepackage{pifont}
\usepackage[ ]{authblk}
\usepackage{chemarrow}
\usepackage{lmodern}
\usepackage{comment}
\usepackage{makecell}
\usepackage{orcidlink}
\UseRawInputEncoding
\hypersetup{
    colorlinks=true,
    linkcolor=blue,
    filecolor=magenta,      
    urlcolor=blue,
    pdftitle={Overleaf Example},
    pdfpagemode=FullScreen,
}
\usepackage{multicol}
\begin{document}
%
%
\title{Comprehensive Dataset and Signal Processing Framework for Phonocardiogram-Based Heart Rate and Blood Pressure Estimation}
%
%

%
%
\author[1,$\ddagger$]{Abdul Ahad Mamun\,\orcidlink{0009-0000-1575-1497}}
\author[1,2,$\ddagger$,*]{Utsab Saha\,\orcidlink{0000-0003-2106-8648}}
\author[1]{Md Hasibul Hasan}
\author[1]{Shahed Ahmed}
\author[1,2]{MD Jahin Alam\,\orcidlink{0000-0002-2922-8696}}

\affil[1]{\small{Department of Electrical and Electronic Engineering\\

Bangladesh University of Engineering and Technology\\
Dhaka 1205, Bangladesh}}
\affil[2]{\small{Department of Computer Science and Engineering\\
BRAC University\\
Dhaka 1212, Bangladesh}}


\affil[$\ddagger$]{\small{\it{These authors contributed equally to this work}}}
\affil[*]{\small{\it{utsab.saha@bracu.ac.bd}}}

%
%

\date{}
\maketitle

\sloppy
%
%
\begin{abstract}
Cardiovascular diseases (CVDs) represent significant global health challenges today, necessitating regular and reliable monitoring to enable early intervention. Phonocardiogram (PCG) signals present a promising non-invasive method for assessing cardiovascular health. While recent studies have focused on estimating heart rate (HR) from PCG signals and blood pressure (BP) through multimodal combinations with other physiological data, reliable and cost-effective systems that can predict both HR and BP using only PCG signals remain largely unexplored. In this study, we proposed and developed a lab-scale cost-effective Phonocardiogram Tracking (PhonoTrack) system that can measure both HR and BP using only the PCG signal. We also introduced a corresponding dataset collected from 15 participants to evaluate the effectiveness of the proposed system. HR was determined using several peak detection methods, such as Hilbert Transform (HT), Shannon Entropy (SE), and WES, achieving notable Pearson correlation coefficients of 0.965, 0.973, and 0.955, respectively. The corresponding root mean square errors (RMSEs) were 2.467 bpm, 1.688 bpm, and 1.992 bpm for HT, SE, and WES, respectively. Additionally, we developed an advanced semi-empirical model based on multiple regression techniques to estimate systolic blood pressure (SBP) and diastolic blood pressure (DBP). This model demonstrated standard deviations of 2.10 mmHg for SBP and 3.20 mmHg for DBP across all subjects, with Pearson correlation coefficients of 0.89 and 0.70, respectively. These findings pave the way for developing a non-invasive, low-cost, and portable PhonoTrack device, positioning it as a promising solution for continuous cardiovascular monitoring settings.

%
\end{abstract}
%
%
%
%

%
%
\section{Introduction}

Real-time health monitoring is critically related to cardiac auscultation, a vital function that plays an essential role in screening for cardiovascular diseases (CVDs) \cite{ahmad2025advancements,mathew2024foundation}. These diseases are often referred to as ``silent killers'' because their symptoms are difficult to detect in early stages \cite{zang2025novel}. According to the World Health Organization (WHO), CVDs are a leading cause of millions of sudden deaths, and their significance in the global health sector is under-recognized due to limited screening, routine diagnostics, and the complexity of clinical management\cite{hanna2002history,gudigar2024automated}. 

Cardiac auscultation involves analyzing the sounds produced as blood circulates through the body, driven by the pumping action of the human heart \cite{pelech2004physiology}. This action generates electrical and mechanical activities, which are represented by the electrocardiogram (ECG) and phonocardiogram (PCG) signals, respectively \cite{ren2018novel}. A typical PCG signal exhibits the first (S1) and second heart sounds (S2) \cite{ren2024comprehensive}. The closure of the mitral (MV) and tricuspid valves (TV) marks the beginning of ventricular systole and produces S1, whereas S2 results from the closure of the aortic (AV) and pulmonary valves (PV), indicating the onset of diastole \cite{ren2024comprehensive,ismail2018localization}. Similarly, other signals such as ballistocardiogram (BCG) and photoplethysmogram (PPG) can also reflect cardiac activity by capturing the mechanical pulse resulting from blood ejection and volumetric changes \cite{shao2016simultaneous,pinheiro2010survey}.

Advanced cardiac imaging techniques, such as computed tomography (CT) and magnetic resonance imaging (CMRI), offer exceptional insights into the detailed structure and function of the cardiovascular system \cite{hussain2022modern}. However, the high cost and operational complexity limit the deployment of these technologies in urgent and resource-constrained healthcare settings \cite{monteiro2023novel}. To address these challenges, researchers have focused extensively on exploring physiological data to develop innovative continuous health monitoring systems over the past few decades \cite{ming2020continuous,gambhir2021continuous,dos2025towards}. By integrating information from ECG and PPG signals, it is possible to estimate heart rate (HR) and blood pressure (BP) using pulse arrival time (PAT) or pulse transit time (PTT) \cite{podaru2020blood,eldakhly2025optimized}. However, a slight misconnection of ECG leads can disrupt alignment with the PPG signal, making it challenging to accurately measure the time difference between the peaks of the PPG wave and the R wave of the ECG, thus reducing measurement accuracy. Furthermore, these signals are insufficient for detecting heart abnormalities, including murmurs and regurgitation.

The incorporation of PCG signal has emerged as a promising solution to overcome these limitations, enabling the development of a cost-effective and reliable continuous health monitoring system \cite{ren2018novel,ismail2018localization,lin2025portable,fynn2025practicality}. Zang et al.~developed a novel non-invasive and wearable device that integrated PCG and ECG signals using ultrasonic PZT transducers and single lead-based sensors for data collection \cite{zang2025novel}. An innovative device consisting of micro-electromechanical (MEM) bionic acoustic sensors and polymer dry electrodes was developed to capture synchronized PCG and ECG signals \cite{monteiro2023novel}. However, the lead wires for the ECG part and their large volume make it unsuitable for wearable and portable applications, ultimately limiting its overall performance and practicality. In addition, measuring the PCG signal with a shielded piezofilm microphone and BCG signal using a weighing scale demonstrated a promising ability to estimate HR and systolic blood pressure (SBP) through signal processing and regression methods based on Random Forest algorithms \cite{gonzalez2022estimation}. Nonetheless, the complexity of BCG data collection and the difficulty in aligning signal peaks from PCG and BCG signals pose challenges in developing a portable and user-friendly device. Additionally, the process of data acquisition faces significant challenges due to its susceptibility to noise and artifacts from various sources, including ambient sounds, physiological interferences, and inconsistent recording conditions \cite{chaudhary2025preprocessing}. Furthermore, the absence of standardized protocols, hardware configurations, and environmental conditions across different studies can result in variations in PCG signal distributions, which may lead to inaccuracy of cardiac performance measurements.

To analyze cardiovascular functionality through feature extraction from PCG signals, several studies have reported datasets considering normal, abnormal, and fetal heart sounds. In 2011, the PASCAL Heart Sound Challenge (HSC) dataset was released to the public, consisting of two subsets: Dataset A and Dataset B \cite{pascal-chsc-2011}. In late 2016, the PhysioNet Computing in Cardiology Challenge database introduced a dataset collected from clinical and non-clinical sources \cite{clifford2016classification}. This dataset comprised a training set labeled from `A' to `F' with 300 valid heart sound recordings. Additional public databases included the Shiraz University Fetal Heart Sounds Database (SUFHSDB), E-General Medical, Michigan Heart Sound and Murmur Database, and the Yaseen and Kwon 2018 dataset \cite{kazemnejad2024open,ismail2018localization,gudigar2024automated}. A few databases have been developed to record ECG and PCG signals simultaneously, allowing us to investigate the synergetic effects of the heart's electro-mechanical activities. For instance, a synchronized acquisition system recorded ECG and PCG signals from 15 adult individuals, enabling precise detection of S1 and S2 heart sounds and accurate heart rate estimation \cite{movahedi2023hardware}. Kazemnejad et al.~established an EPHNOGRAM dataset, which included simultaneous ECG and PCG data from 24 healthy adults in resting and stress test scenarios \cite{kazemnejad2024open}. Their work aimed to develop a robust hardware-software prototype capable of classifying physical activities and analyzing electro-mechanical dynamics by measuring key features, such as the R-peaks in the ECG, the S1 and S2 components of the PCG, QRS duration, time differences between R--R, and S1--S2 intervals. However, the size of the prototype presented challenges in continuous data collection. 

Recently, researchers have made significant advances in PCG signals, focusing on investigating HR and cardiac abnormalities using support vector machines (SVMs), hidden Markov models (HMMs), and advanced machine learning (ML) techniques \cite{gudigar2024automated,touahria2023feature,chen2017calculating}. However, these methods are often complex, expensive, and unsuitable for portable devices \cite{hamza2024comprehensive}. Additionally, these studies have primarily focused on estimating individual cardiac metrics, along with facing difficulties in predicting HR and BP simultaneously from PCG signals. An alternative approach involves using a low-cost acoustic sensor and a signal processing method to enable the creation of a simple and lightweight system that can serve as a wearable device for continuous health monitoring \cite{sa2012low,angelucci2025wearable}. Moreover, measuring crucial cardiac metrics like BP alongside HR using only the PCG signal presents a promising strategy for developing a novel phonocardiogram tracking (PhonoTrack) device. Thus, there are exciting opportunities to explore new setups for collecting PCG signals and analyzing them with simple algorithmic strategies to determine HR and BP, leading to the development of an effective and efficient PhonoTrack device.

In this work, we developed a low-cost and effective data acquisition setup to capture PCG and ECG signals simultaneously, with the aim of advancing a portable and wearable PhonoTrack device. Beyond the hardware prototype, we present an end-to-end and reproducible framework that links synchronized acquisition, signal-quality assessment, and frame-wise algorithmic evaluation for PCG-only cardiovascular monitoring. We introduce a new database comprising synchronized PCG and ECG recordings along with BP data, collected from 15 adults over 60-second intervals. To our knowledge, this is one of the few datasets that provides time-aligned PCG-ECG recordings with paired cuff BP in the same session, enabling direct benchmarking of PCG-derived HR and BP against reference measurements. The ECG and BP signals were used as reference standards for validating our system. To analyze and validate the PCG data, we applied multiple signal processing techniques, including fast Fourier transform (FFT), spectrogram analysis, mel-frequency cepstral coefficients (MFCCs), and wavelet energy spectrum (WES), to extract frequency-domain features and distinguish key heart sounds such as S1 and S2. Importantly, we use these analyses not as isolated visualizations, but as a systematic signal-quality and interpretability layer that supports reliable S1/S2 characterization in short windows. Normalized root mean square error (NRMSE) between WES and traditional energy spectrum (ES) representations, using several wavelets such as Morlet, Morse, and Bump, was used to assess signal quality. We further report quality statistics across subjects to make the dataset's recording conditions and variability transparent. Three HR estimation algorithms -- based on Hilbert Transform (HT), Shannon Entropy (SE), and wavelet-based peak detection -- were employed and compared against the ECG reference signal to evaluate consistency. To more effectively evaluate the variability among different methods, we conducted Bland-Altman analyses for each subject. For BP estimation, a semi-empirical model was developed using HR and PCG-derived features and validated by calculating standard error metrics. The extracted features included the S1--S2 and S2--S1 intervals, as well as the rising and falling slopes of S1 and S2, along with their respective durations. To reduce overfitting risk under small-sample regression, we additionally evaluate BP estimation under subject-wise leave-one-subject-out cross-validation (LOOCV), reporting out-of-sample MAE and RMSE along with Bland-Altman agreement in the supplementary material. 

This study mostly focuses on the integration and validation of hardware components and standard digital signal processing (DSP) techniques within the proposed PhonoTrack system. The system employs well-known signal processing methods in a unified framework to simultaneously estimate HR and BP using PCG signals, indicating that standard DSP techniques are more applicable and efficient than more complex alternatives. The primary contribution of this research is the demonstration of the system's feasibility, reliability, and cost-effectiveness through experimental evaluation. Overall, the study highlights the implementation and real-world validation of the proposed PhonoTrack system, laying the foundation for a non-invasive, low-cost continuous cardiovascular monitoring system based on PCG signals.



%
%
\section{Methodology}
\subsection{Dataset Collection}
Our proposed dataset includes 15 synchronized ECG and PCG recordings, as well as the corresponding BP data from the volunteers. The ECG was captured using a three-lead, single-channel setup, while the PCG signal was obtained using our custom-designed PhonoTrack device. The ECG-derived HR served as the gold standard for validating the PCG-based HR estimation, and BP readings taken via a manual sphygmomanometer were used as the reference values for BP estimation. This study included only male participants aged between 18 and 50 years, with data collected from 15 healthy volunteers. All participants were free from any known cardiovascular diseases and provided informed consent before the recordings. We excluded children and individuals with diagnosed cardiac conditions, implanted cardiac devices, and chest issues that could affect sensor placement and signal quality. Data acquisition was repeated if significant noise interference and sensor displacement were observed, ensuring the quality of the signals. All recordings were conducted under the supervision of a trained medical professional. Each recording session lasted for 60 seconds, during which PCG and ECG signals were acquired simultaneously.

To minimize noise and motion-related artifacts during data acquisition, all recordings were conducted in a controlled indoor laboratory environment. Participants were comfortably seated in an armchair to maintain consistency and eliminate variability due to physical activity related to body movement and posture. The diaphragm for PCG signal recording was placed on the chest with a comfortable and adjustable strap, which provided consistent contact between the sensor and the skin throughout the recording session. This strap-based setup helped minimize relative motion between the sensor and the body, effectively reducing motion-induced artifacts and signal fluctuations. Additionally, environmental noise was minimized by conducting recordings in a quiet setting and ensuring stable sensor placement before data collection. Collectively, these strategies enhanced signal stability, reliability, and overall data quality during PCG acquisition.

%
%
\begin{figure}[hbt]
    \centering
    \includegraphics[width =0.99\linewidth, center]{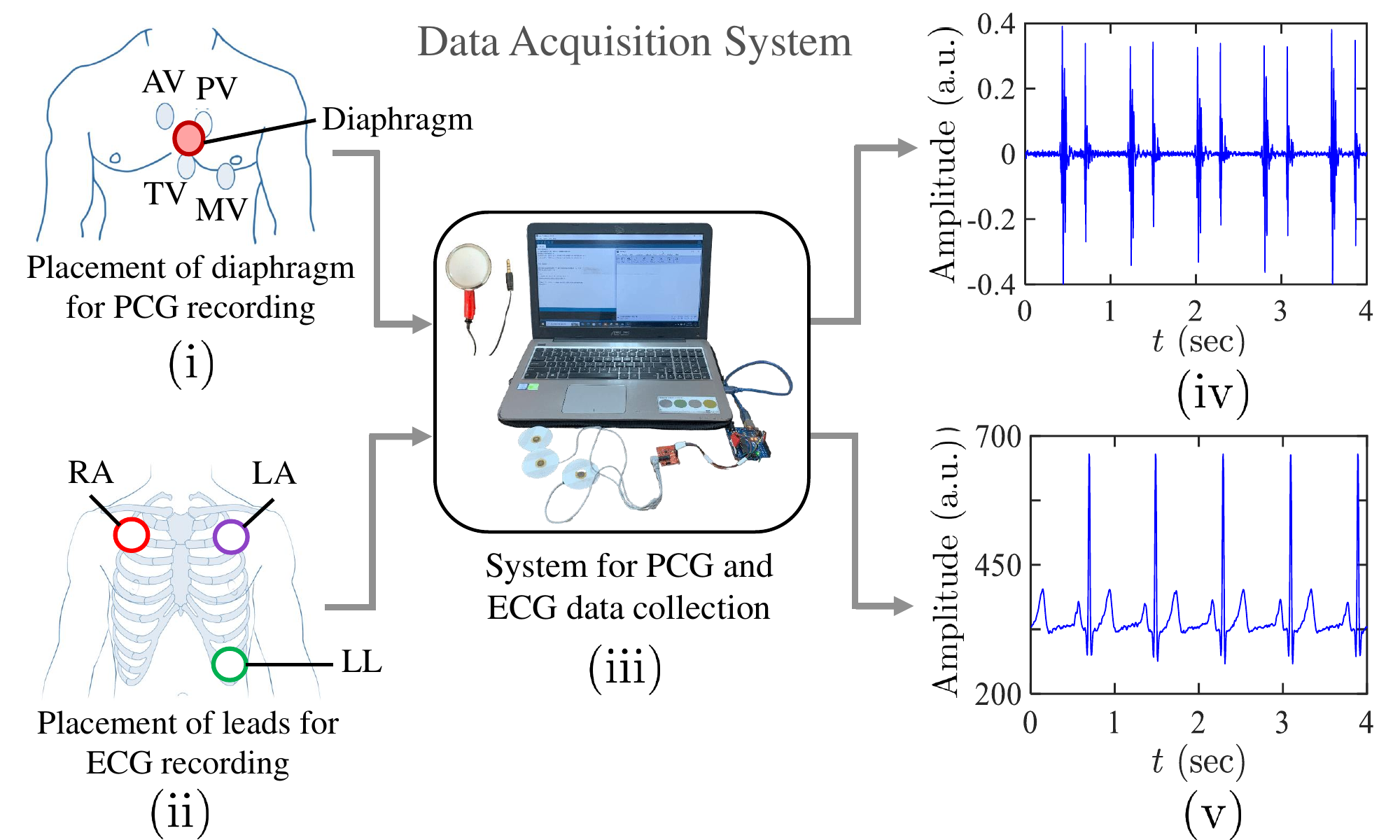}    
    \caption{Overview of the Data Acquisition System: (i) Diaphragm placement for phonocardiogram (PCG) signal recording, where the red circle indicates the position of the diaphragm for our developed novel phonocardiogram tracking (PhonoTrack) device; (ii) Leads placement for the three-lead electrocardiogram (ECG) system, following the standard configuration for accurate signal acquisition; (iii) PCG and ECG signals recording system; (iv) Collected PCG signal; (v) Collected ECG signal.}
    \label{Fig1M}
\end{figure}
%

Prior to the signal acquisition, the participant's blood pressure was measured manually using a sphygmomanometer. Immediately following the 60-second recording session, BP was measured again. The average of the two BP readings was considered the participant's ground truth BP for that session. It should be noted that arterial pressure varies across minutes. Clinical guidelines recommend taking repeated BP measurements 1--2 minutes apart and averaging them to obtain a more representative value~\cite{mancia20132013, shimamoto2014japanese, chadachan2018understanding}. We therefore used the average of the manual BP readings taken immediately before and after the 60-s PCG segment as a pragmatic estimate of the subject's mean BP during that window, following the guidelines provided by the American Heart Association (AHA)~\cite{AHA_MeasureBP_Accurately}. All sessions were conducted under the supervision of a trained medical professional to ensure participant safety and procedural accuracy. The complete data acquisition setup, including the PCG and ECG recording configuration and representative sample waveforms, is illustrated in Fig.~\ref{Fig1M}.

The custom-built PhonoTrack device was designed to ensure high-quality heart sound recording while maintaining a cost-effective setup for PCG acquisition. To achieve optimal signal clarity, a microphone was embedded inside the extended tube of a standard stethoscope, allowing it to capture heart sounds directly from the chest with minimal external noise interference. The acoustic signals collected through the stethoscope's diaphragm were transmitted via a wired connection to an iPad sound recorder, ensuring minimal signal loss and real-time data acquisition. This setup effectively preserved the natural frequency characteristics of heart sounds, providing a reliable method for phonocardiographic analysis. The iPad's built-in audio recording application was used to record and store the PCG data in a digital format, facilitating easy playback, signal processing, and further analysis. By default, the iPad records audio at 44.1~kHz, which is substantially higher than the dominant spectral content of PCG signals. Therefore, in addition to the original recordings, we provide a down-sampled PCG version at 2~kHz to reduce storage and computational cost. Prior to downsampling, the signals are low-pass filtered (anti-aliasing) to suppress frequency components above the new Nyquist frequency, ensuring that the clinically relevant PCG band is preserved while avoiding aliasing artifacts. For consistency in processing, all PCG recordings were decoded and stored as 16-bit PCM prior to analysis, corresponding to a 16-bit quantization resolution for the waveform samples.

It should be noted that cardiac auscultation is typically performed in four main areas of the chest~\cite{Dornbush}. The sound of the TV is best heard on the left side of the heart, specifically between the fourth and sixth ribs \cite{liu2016open,oliveira2021circor}. However, we opted to place the diaphragm of the proposed PhonoTrack device slightly higher than the TV position to collect the PCG signal. The PhonoTrack system was positioned at a standardized anatomical site for all subjects, specifically in the left parasternal region, one intercostal space above the typical location for the tricuspid valve site, identified through rib palpation. Recordings were made by the same trained operator, who assured consistent device orientation and contact pressure. Subjects were seated upright in a relaxed posture and instructed to breathe normally, avoiding speaking and taking deep breaths during the recording process. This adjustment in diaphragm placement ensures high-quality recordings of the S1 and S2 heart sounds while maintaining controlled conditions concerning respiration and posture.

The placement of the diaphragm while collecting PCG is illustrated in Fig.~\ref{Fig1M}(i). As previously mentioned, the ECG acquisition system is designed using a three-lead ECG configuration, where electrodes are strategically placed to capture the heart's electrical activity, as illustrated in Fig.~\ref{Fig1M}(ii). The ECG leads are positioned on the two hands and the lower left rib or left leg (LL), following the standard three-lead ECG configuration. The leads are connected to an ECG sensor module (AD8232), which serves to amplify weak cardiac signals while effectively filtering out noise using a high-pass filter. The AD8232 incorporates a double-pole high-pass filter designed to eliminate motion artifacts and electrode half-cell potential. Additionally, it features an operational amplifier that utilizes a three-pole low-pass filter, allowing for the removal of extraneous noise. The collected ECG signal is digitized using an analog-to-digital converter (ADC) and transmitted to a custom-built Arduino-based acquisition system. The Arduino collects the digitized ECG data and transfers it in real time via USB to a laptop, where it is stored in text format for further analysis. The equipment setup for data acquisition and real-time simultaneous data collection is illustrated in Figs.~S1 and S2 of the supplementary materials. The PCG signal from the iPad and the ECG stream from the Arduino were recorded manually using separate buttons. Although these datasets were gathered almost simultaneously, slight time mismatches may have occurred. To address this issue, we corrected any resulting latency during the post-processing phase by utilizing the time indices of each signal. Additionally, we verified and improved the alignment by focusing on primary physiological features, such as the R-peaks in the ECG and the corresponding S1 heart sounds. This approach ensured precise temporal correspondence between the two streams for further analysis.


%
\subsection{Signal Processing} 
\subsubsection{Heart rate estimation}

The collected ECG data was sampled at an average frequency of $f_{s,\text{ecg}} = 190$ Hz. This sampling rate is adequate for R-peak detection because the QRS complex is a relatively short, high-slope event (typically on the order of $\sim$80--120 ms~\cite{noble1990electrocardiography, surawicz2009aha}), so $190$ Hz provides about 15--23 samples across a QRS complex and is sufficient to localize the R peak with standard bandpass filtering and peak-picking. The process of estimating HR from ECG was initiated by segmenting the ECG signal into non-overlapping frames of 6-second duration. To mitigate baseline drift and other low-frequency artifacts, we applied a detrending operation to each ECG frame. We then normalized the detrended ECG signal by its maximum absolute amplitude to improve threshold-based peak detection. R-peaks are the most prominent features within the QRS complex of an ECG signal, typically showing a relative prominence of at least 0.7 when the waveform is normalized \cite{rabbani2011r}. In wavelet-based and multiscale analyses of ECG data, peak detection thresholds are usually specified as fractions of the signal's maximum value or based on local statistical measures \cite{farrokhi2025reliable}. These thresholds are adjusted empirically to achieve an optimal balance between sensitivity and specificity. To identify the R-peaks in the ECG, we employed a local maxima detection algorithm and set a prominence threshold of 0.8 to ensure accurate and reliable detection across all samples. We extracted the temporal locations of these peaks and computed the average inter-peak interval for each frame. Finally, the HR was calculated using the following equation: 
\begin{equation}
\label{eq:HR}
    \text{HR}_{\text{ecg}} = \frac{60 \times f_{s,\text{ecg}}}{\text{avg}(\Delta P)},
\end{equation}
where $\Delta P$ represents the intervals between detected R-peak indices in a 6-sec frame. 

To estimate HR from PCG, we utilized a 6-second PCG frame synchronized with the ECG frame. The collected PCG signal was sampled at a frequency of \( f_{s,\text{pcg}} = 44.1 \, \text{kHz} \). The PCG signal was processed in a frame-by-frame manner to ensure consistency and accuracy in heart sound analysis. Each frame underwent a wavelet-based denoising process to enhance signal quality by reducing high-frequency noise. Specifically, we employed a Daubechies wavelet (db4) with eight levels of decomposition to decompose the raw PCG signal into multiple frequency components. This wavelet shrinkage denoising method aims to reduce noise by eliminating wavelet coefficients with small magnitudes that are primarily influenced by noise, while preserving larger coefficients that represent significant signal structures. Recent studies have set thresholds as a percentage of the maximum signal amplitude, typically up to 20\%, in PCG analysis to achieve a balance between noise suppression and signal preservation \cite{ergen2013comparison}. In our approach, we employed a thresholding-based shrinking method with the noise level set at 15\% of the maximum signal amplitude, effectively minimizing unwanted noise.

\begin{figure}[hbt]
    \centering
    \includegraphics[width =1.0\linewidth, center]{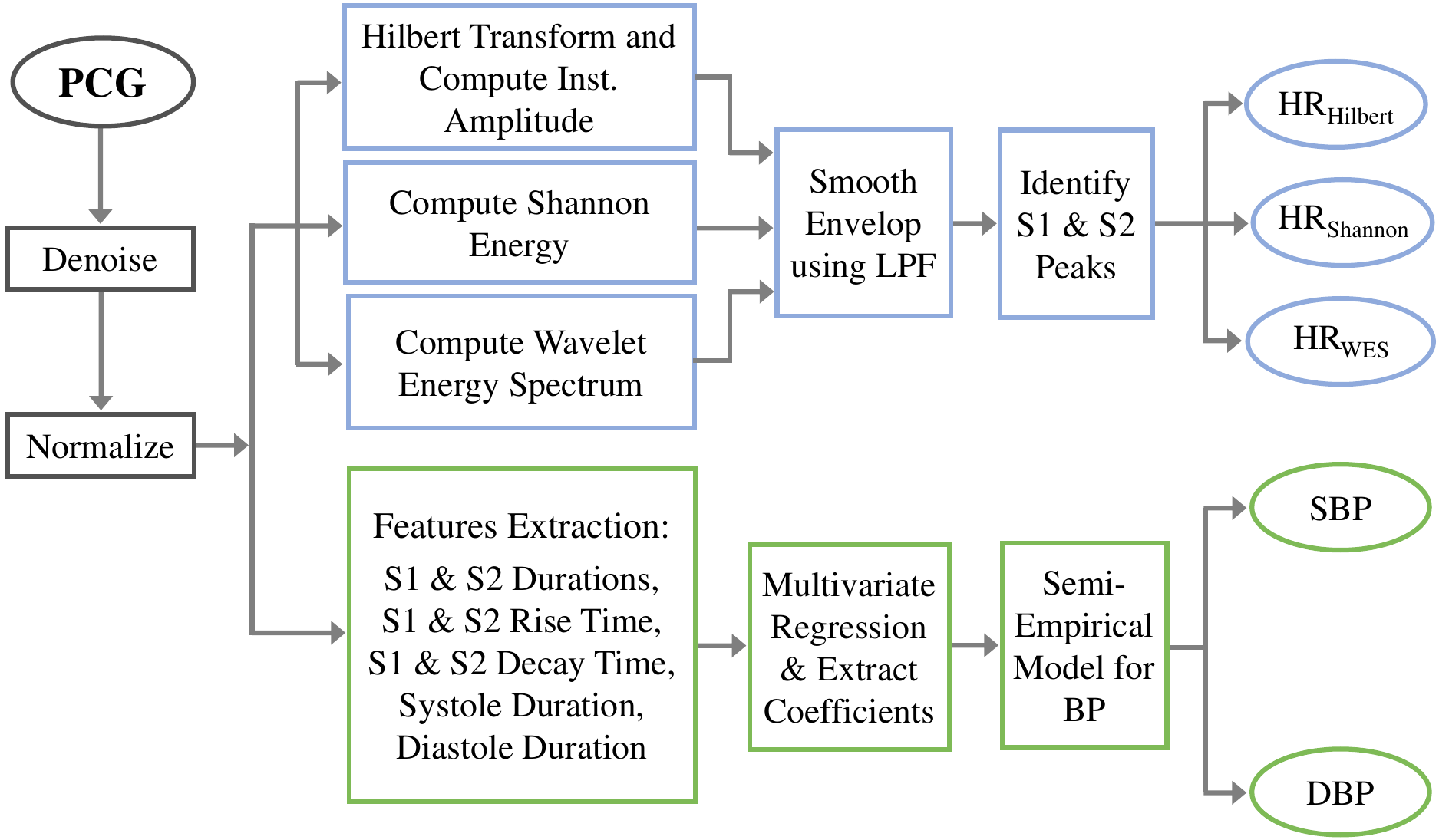}
    \caption{Flowchart of the proposed signal processing-based algorithms for estimating heart rate (HR) and blood pressure (BP).}
    \label{Fig2M}
\end{figure}

Finally, the denoised signal was reconstructed to obtain a cleaner version of the original PCG frame. Following the denoising step, each frame was normalized by dividing the signal by its maximum value. This normalization process ensured consistency across different frames and mitigated amplitude variations due to recording conditions. To estimate heart rate from a normalized PCG frame, it was necessary to extract the envelope of S1 and S2 from the PCG signal. This extraction was achieved using three different signal processing techniques as follows: (i) HT method; (ii) SE function approach; and (iii) WES-based envelope detection procedure. We selected these envelope-based methods because HR estimation from PCG ultimately depends on reliably detecting successive heart-sound events (S1 or S2) within short windows, since the S1--S1 (or S2--S2) interval directly corresponds to the cardiac cycle duration. The Hilbert envelope is a simple time-domain amplitude envelope that highlights burst-like heart-sound activity and is commonly used as a lightweight front-end for event detection in PCG-style segmentation pipelines~\cite{prasad2020detection}. Shannon-energy-based envelopes have been widely used to sharpen transient components and support peak-based S1/S2 segmentation by emphasizing higher-energy events relative to background fluctuations~\cite{arjoune2024noise}. Finally, wavelet energy spectrum is used to localize concentrated energy ridges in the time--frequency plane and to improve peak saliency and robustness under different noise~\cite{yamacli2008segmentation}. The flowchart of the HR estimation process from PCG is depicted in Fig.~\ref{Fig2M}. 

%
\textbf{Hilbert Transform-based envelope detection:} HT is a signal processing technique used to extract instantaneous characteristics of a signal, including amplitude and phase. It is particularly useful for analyzing non-stationary signals such as heart sounds \cite{atbi2013separation}. In this study, we employed the HT to obtain the envelope of a PCG frame, ${x}[n]$. By applying the HT, we generated a complex-valued version of the signal, $z[n]$, where the absolute value, $|z[n]|$, corresponds to the instantaneous amplitude or envelope. The envelope of the PCG frame was detected as follows \cite{benitez2001use}: 
\begin{subequations}
\begin{equation}
\label{eq:hil1}
\hat{x}[n] = \sum_{k=-\infty}^{\infty} x[k] \cdot h[n - k], \quad \text{where} \quad
h[n] =
\begin{cases}
\displaystyle\frac{2}{\pi n}, & \text{if } n \text{ is odd} \\
0, & \text{if } n \text{ is even}
\end{cases}
\end{equation}
\begin{equation}
\label{eq:hil2}
e[n] = |z[n]| = \sqrt{x[n]^2 + \hat{x}[n]^2}, \quad \text{where} \quad z[n] = x[n] + j \cdot \hat{x}[n].
\end{equation}
\end{subequations}
Note that the Hilbert envelope was computed from the analytic signal (FFT-based implementation in MATLAB, \texttt{hilbert}).

%
%

%
\textbf{Shannon energy-based envelope detection:} The second method for extracting the PCG envelope was based on the SE function, which emphasizes high-energy regions while suppressing low-energy components. This method enhances significant signal features, making it particularly useful for detecting the envelope of heart sounds. After normalizing the denoised PCG signal, we computed the Shannon energy envelope, $E_{\rm SE}[n]$, using the following expression \cite{beyramienanlou2017shannon}:
\begin{equation}
\label{eq:shan}
E_{\rm SE}[n] = -x[n]^2 \cdot \log\left(x[n]^2+\epsilon\right),
\end{equation}
which effectively captured the dominant peaks corresponding to S1 and S2. In this formulation $\epsilon$ is a small constant (e.g., $10^{-12}$) added for numerical stability to avoid $\log(0)$. Additionally, a zero-phase Butterworth low-pass filter was implemented via forward–backward filtering (\texttt{filtfilt}) for edge handling.

\textbf{Wavelet energy spectrum-based envelope detection:} The third method leveraged the WES to analyze the PCG signal across multiple frequency bands. The process began with a wavelet transform, which decomposed the PCG signal into different frequency components using the following expression \cite{Ergen01042012}:
\begin{equation}
\label{ew:CWT}
    \text{CWT}(n) = \frac{1}{\sqrt{\left| s \right|}}\sum_{i=n-\frac{M}{2}}^{n+\frac{M}{2}}x(i)\psi(\frac{i-n}{s}),
\end{equation}
where $\psi$ represents the mother wavelet function, while $x$ denotes the PCG frame. The variables $n$ and $s$ refer to the dilation and translation parameters, respectively. Additionally, $M$ is the duration of the wavelet when $s=1$. The term $\frac{1}{\sqrt{\left| s \right|}}$ is included to ensure the preservation of energy.
This decomposition allowed us to extract the fine details of the heart sounds, particularly the S1 and S2 peaks. To construct the WES envelope, we squared the absolute values of the wavelet coefficients, obtaining the wavelet energy spectrum. To compute a global energy distribution, we summed the values of wavelet energy across all decomposition levels by using the following equation \cite{Ergen01042012}: 
\begin{equation}
\label{eq:WES}
    \text{WES}(n) = \frac{1}{N}\sum_{k=1}^{L}\left| \text{CWT(n,k)} \right|^2,
\end{equation}
where $L$ represents the number of scales and $N$ denotes the length of the PCG frame. This process resulted in a smooth and robust envelope representation, effectively highlighting key features of the PCG signal while suppressing noise and irrelevant fluctuations. A detailed flowchart summarizing the full signal-processing pipeline (envelope extraction with HT, Shannon and WES) is provided in the supplementary material (Fig.~S9).

After obtaining the envelope, it was then smoothed using a low-pass Butterworth filter with a cutoff frequency of 20 Hz to remove high-frequency noise. Although envelope-based PCG segmentation often applies a lower cutoff (e.g., the widely used Schmidt/Springer homomorphic-envelope implementation uses an 8 Hz low-pass by default~\cite{springer2015logistic}), using such a low cutoff in our setting over-smooths the short S1 or S2 bursts and degrades peak localization for several envelope detection algorithms. This is confirmed by the added sensitivity analysis (Figure S9 of the supplementary materials): at 8 Hz, the HR MAE becomes very large, whereas for cutoffs ≥10 Hz the MAE remains consistently low, with a broad stable region around 15--30 Hz. Therefore, we retained 20 Hz as a conservative choice that preserves the localization of S1 and S2 sounds while still providing sufficient smoothing. The next step was to detect the S1 and S2 peaks, which were achieved by identifying the local maxima in the smoothed envelope of the PCG signal. 

\begin{figure}[hbt]
    \centering
    \includegraphics[width =0.90\linewidth, center]{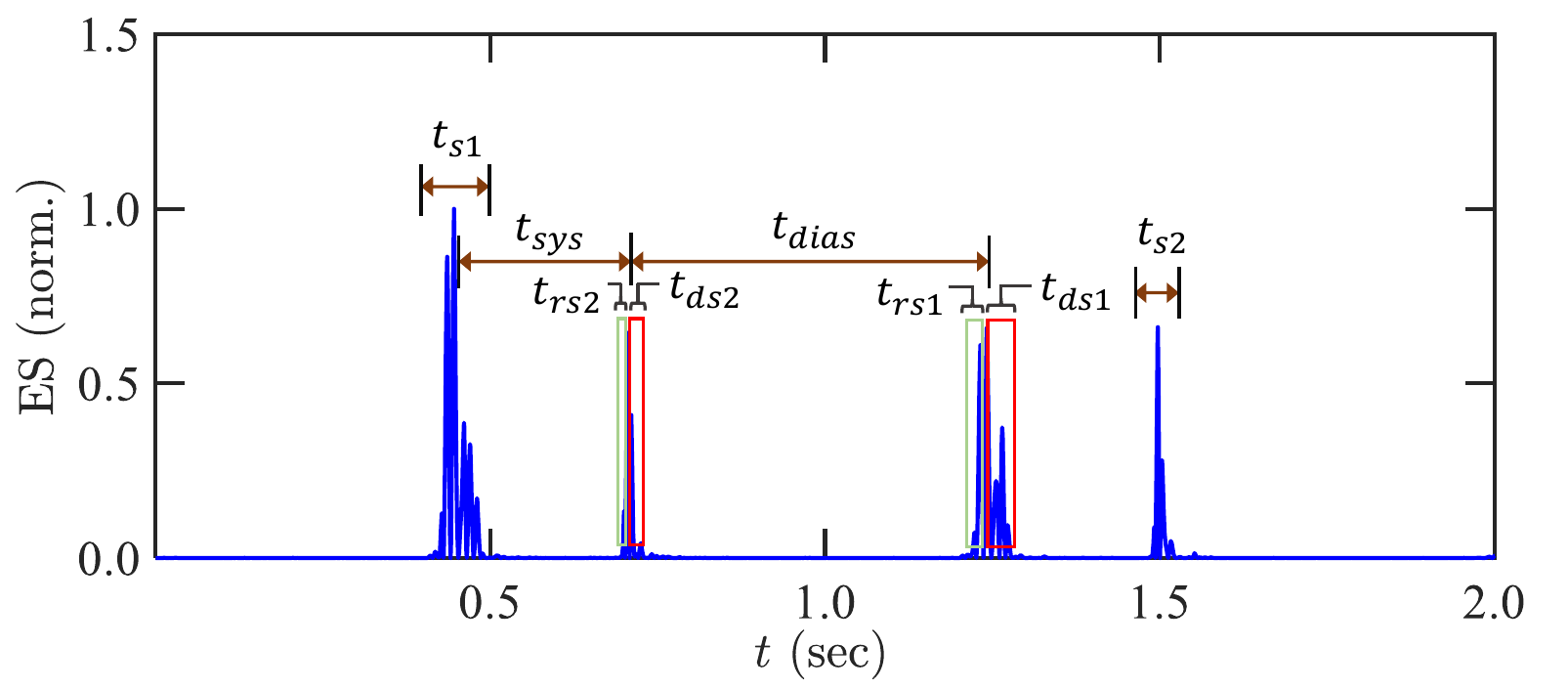}
    \caption{Energy spectrum (ES) of a 2-sec PCG frame showing different features for blood pressure (BP) estimation.}
    \label{Fig3M}
\end{figure}

The method for detecting the S1 and S2 heart sounds involved identifying local maxima in the envelope of the phonocardiogram. To ensure the realistic intervals between the peaks, we applied a minimum peak distance of 0.125 sec between S1 sounds and between S2 sounds, as a healthy human heart rate cannot exceed 240 beats per minute. This approach is consistent with previous heart sound segmentation methods, which incorporated temporal constraints reflecting the dynamics of the cardiac cycle to avoid false peak detection \cite{liu2018automatic,strazza2018pcg}. Additionally, we set a minimum peak height threshold at 15\% of the maximum absolute envelope value to filter out low-energy fluctuations and noise. This technique aligns with findings from several studies that indicate relatively low amplitude thresholds are effective in heart sound detection, enabling the reliable distinction of significant heart sound features from background noise \cite{springer2015logistic}. After identifying the local maxima, we proceeded to determine the S1 and S2 peaks by analyzing the time intervals between them. If the time difference between the first two peaks is smaller than the time difference between the second and third peaks, the first two peaks are alternately assigned to S1 and S2. To facilitate peak detection benchmarking, manual ground-truth S1 and S2 peak annotations are additionally provided for a representative subset of recordings (Supplementary Fig.~S5--S7), along with the corresponding subject metadata (age, sex, height, weight, and reference BP) reported in the figure captions. Once the S1 and S2 peaks were identified, we calculated the average of the systolic, $t_{\text{sys}}$, and diastolic durations, $t_{\text{dias}}$. The distance between S1 and S2 peaks is defined as $t_{\text{sys}}$, and the distance between S2 and S1 peaks is considered $t_{\text{dias}}$, as shown in Fig.~\ref{Fig3M}. The corresponding expressions can be written as follows:  
\begin{subequations}
\begin{equation}
\label{eq:SAD}
    t_{\text{sys}} = \frac{1}{N_C} \sum_{i=1}^{N_C} \frac{|s_i - e_i|}{f_{s,\text{pcg}}},
\end{equation}
\begin{equation}
\label{eq:DAD}
    t_{\text{dias}} = \frac{1}{N_C} \sum_{i=1}^{N_C} \frac{|e_i - s_{i+1}|}{f_{s,\text{pcg}}},
\end{equation}
\end{subequations}
where $s_i$ and $e_i$ are the start of systole and diastole, respectively, where $i$ is the index of the sample point for the PCG signal. $N_C$ is the total number of cardiac cycles. It is important to note that we computed HR using the systolic and diastolic intervals individually. To obtain the overall HR$_{\rm pcg}$, we averaged the values calculated from the systolic and diastolic intervals using the following expression:
\begin{equation} 
\label{eq:HR_pcg}\text{HR}_{\text{pcg}}= \left( \frac{1}{t_\text{sys}}+ \frac{1}{t_\text{dias}} \right) \times \frac{60}{2}.
\end{equation}

%
%

%
\subsubsection{Blood pressure estimation}
For BP estimation, we focus on physiologically interpretable timing features derived from the S1/S2 dynamics, since these time-interval descriptors reflect cardiac mechanical timing and are influenced by vascular loading. Accordingly, we use a lightweight semi-empirical regression model that maps the PCG-derived timing features (together with $\mathrm{HR}_{\mathrm{pcg}}$) to SBP/DBP, which keeps the estimator suitable for wearable implementation while remaining transparent for analysis and ablation. In this model, the input features are extracted directly from each PCG frame as described next.
These features include the durations of S1 and S2, the rise times of S1 and S2, the decay times of S1 and S2, $t_{\text{sys}}$, $t_{\text{dias}}$ and $\text{HR}_{\text{pcg}}$. We extracted these features from each frame of the PCG signal. We previously described the process of detecting S1 and S2 peaks in detail, as well as the process of extracting $t_{\text{sys}}$, $t_{\text{dias}}$ and $\text{HR}_{\text{pcg}}$ features. For each detected S1 peak, we calculated the rise and decay times by tracking the duration it takes for the envelope amplitude to exceed or fall below a small threshold. The time taken for the envelope to increase from a low value to reach the S1 peak is referred to as the systolic rise time ($t_{\rm rs1}$), as shown in Fig.~\ref{Fig3M}. Conversely, the time taken for the envelope to decrease from the S1 peak back down to the baseline is called the systolic decay time ($t_{\rm ds1}$). We computed these features using the following equations:
\begin{subequations}
\begin{equation}
\label{eq:rs1}
    t_{\text{rs1}} = \frac{1}{N_C} \sum_{i=1}^{N_C} \frac{r_{i,s1}}{f_{s,\text{pcg}}},
\end{equation}

\begin{equation}
\label{eq:rs2}
    t_{\text{ds1}} = \frac{1}{N_C} \sum_{i=1}^{N_C} \frac{d_{i,s1}}{f_{s,\text{pcg}}},
\end{equation}
\end{subequations}
where $r_{i,s1}$ represents the rise time index for the $i^\text{th}$ cardiac cycle, while $d_{i,s1}$ denotes the decay time index for the same cycle. For each detected S2 peak, the diastolic rise ($t_{\rm rd2}$) and decay times ($t_{\rm dd2}$) were determined in the same manner as for the systolic peaks using the following expressions:
\begin{subequations}
\begin{align}
    t_\text{rd2} = \frac{1}{N_C} \sum_{i=1}^{N_C} \frac{r_{i,d2}}{f_{s,\text{pcg}}},
\end{align}
\begin{align}
    t_\text{dd2} = \frac{1}{N} \sum_{i=1}^{N_C} \frac{d_{i,d2}}{f_{s,\text{pcg}}}.
\end{align}
\end{subequations}
where $r_{i,d2}$ represents the rise time index for the $i^\text{th}$ cardiac cycle, while $d_{i,d2}$ denotes the decay time index for the same cycle.

%
\begin{table}[hbt]
\centering
\caption{Summary of the coefficients extracted using multiple regression for estimating systolic (SBP) and diastolic blood pressures (DBP) from phonocardiogram (PCG) signal.}
\label{Table1M} 
\begin{tabular}{l p{3cm} l p{3cm} p{2.5cm}} 
\Xhline{3\arrayrulewidth}
{Name} & {Value} & {Name} & {Value} & {Unit} \\ [0.5ex] 
\hline
$C_1$ & $6.55 \times 10^{-1}$  & $C_2$ & $1.799$ & mmHg \\
$\sigma_{\text{sys}}$ & $-1.112$  & $\alpha_{\text{dias}}$ & 2.463 & mmHg$\cdot$ms$^{-1}$ \\
$\sigma_{\text{rs1}}$ & $4.890 \times 10^1$  & $\alpha_{\text{rd2}}$ & $1.196 \times 10^1$ & mmHg$\cdot$ms$^{-1}$ \\
$\sigma_{\text{rs1}}$ & $4.845 \times 10^1$  & $\alpha_{\text{dd2}}$ & $7.878$ & mmHg$\cdot$ms$^{-1}$ \\
$\sigma_{\text{s1}}$ & $-5.044 \times 10^1$  & $\alpha_{\text{s2}}$ & $-9.643$ & mmHg$\cdot$ms$^{-1}$ \\
$\sigma_{\rm{s,HR}}$ & $-6.940 \times 10^{-3}$  & $\alpha_{\rm{d,HR}}$ & $-5.291 \times 10^{-2}$ & mmHg$\cdot$(bpm)$^{-1}$ \\
$\sigma_{\rm{s,{HR}^2}}$ & $0.41$ $\times$ 10$^{-5}$  & $\alpha_{\rm{d,{HR}^2}}$ & $3.05$ $\times$ 10$^{-4}$ & mmHg$\cdot$(bpm)$^{-2}$ \\
$\sigma_{\text{sys}^2}$ & $1.689$  & $\alpha_{\text{dias}^2}$ & $-2.816$ & mmHg$\cdot$s$^{-2}$ \\
$\sigma_{\text{s1}^2}$ & $2.298 \times 10^1$  & $\alpha_{\text{s2}^2}$ & $3.049 \times 10^1$ & mmHg$\cdot$s$^{-2}$ \\
$\sigma_{\text{rds1}}$ & $-5.902 \times 10^1$ & $\alpha_{\text{rdd2}}$ & $-1.094 \times 10^1$ & mmHg$\cdot$s$^{-2}$ \\
\Xhline{3\arrayrulewidth}
\end{tabular}
\end{table}
%

The next feature was the peak duration of S1, $t_\text{s1}$, which was simply the sum of $t_{\text{rs1}}$ and $t_{\text{ds1}}$. Similarly, we calculated the peak duration of S2, $t_\text{ts2}$. All of these features are pictorially described in Fig.~\ref{Fig3M}, where we show the corresponding time duration of each feature. Finally, we determined the average HR of each subject by averaging their corresponding frames' HR. Note that after determining the feature values for all the frames, we calculate the average, which represents the respective average features of a subject. Following the feature extraction step, we developed a semi-empirical model to determine SBP and DBP from various features extracted from PCG signals. We established a relationship between BP and eight features extracted from the PCG, which proved to be very useful for our analysis. The value of SBP, denoted as \(P_\text{sys}\), and value of DBP, denoted as \(P_\text{dias}\), are determined using the following equations:

\begin{subequations}
\begin{equation}
\label{eq:Psys}
\begin{aligned}
P_{\text{sys}} = & \, C_1 + \sigma_{\text{sys}} \cdot t_{\text{sys}} + \sigma_{\text{rs1}} \cdot t_{\text{rs1}} + \sigma_{\text{ds1}} \cdot t_{\text{ds1}} \\
& + \sigma_{\text{s1}} \cdot t_{\text{s1}} + \sigma_{\rm{s,HR}} \cdot \text{HR}_\text{pcg} + \sigma_{\rm{s,{HR}^2}} \cdot \text{HR}_\text{pcg}^2 + \sigma_{\text{sys}^2} \cdot t^2_{\text{sys}} \\
& + \sigma_{\text{s1}^2} \cdot t^2_{\text{s1}} + \sigma_{\text{rds1}} \cdot t_{\text{rs1}} t_{\text{ds1}},
\end{aligned}
\end{equation}
\begin{equation}
\label{Eq:pdias}
\begin{aligned}
P_{\text{dias}} = & \, C_2 + \alpha_{\text{dias}} \cdot t_{\text{dias}} + \alpha_{\text{rd2}} \cdot t_{\text{rd2}} + \alpha_{\text{dd2}} \cdot t_{\text{dd2}} \\
& + \alpha_{\text{s2}} \cdot t_{\text{s2}} + \alpha_{\rm{d,HR}} \cdot \text{HR}_\text{pcg} + \alpha_{\rm{d,{HR}^2}} \cdot \text{HR}_{\text{pcg}}^2 + \alpha_{\text{dias}^2} \cdot t^2_{\text{dias}} \\
& + \alpha_{\text{s2}^2} \cdot t^2_{\text{s2}} + \alpha_{\text{rdd2}} \cdot t_{\text{rd2}} t_{\text{dd2}},
\end{aligned}
\end{equation}
\end{subequations}
The coefficients and parameters associated with the semi-empirical model are summarized in Table \ref{Table1M}. These coefficients were obtained through multiple regression analyses~\cite{cohen2013applied}. The entire process of calculating SBP and DBP is illustrated in Fig.~\ref{Fig2M}. It should be noted that, given the limited effective sample size of 15 subjects, the proposed semi-empirical regression model should be interpreted cautiously. The inclusion of multiple linear, quadratic, and interaction terms under such a small-data setting increases the risk of overfitting and results in coefficient instability (refer to the supplementary material, Table S3). Therefore, the coefficients derived from this model should not be considered definitive population-level physiological parameters, but rather as preliminary estimates obtained from this pilot dataset collected using the PhonoTrack system. Consequently, the current semi-empirical model should be viewed as an initial proof-of-concept baseline. Future studies with larger and more diverse cohorts will be essential to enhance model stability, reduce overfitting, and establish broader generalizability.

%
%
\section{Data Validation}

\begin{figure}[hbt]
    \centering
    \includegraphics[width =1.0\linewidth, center]{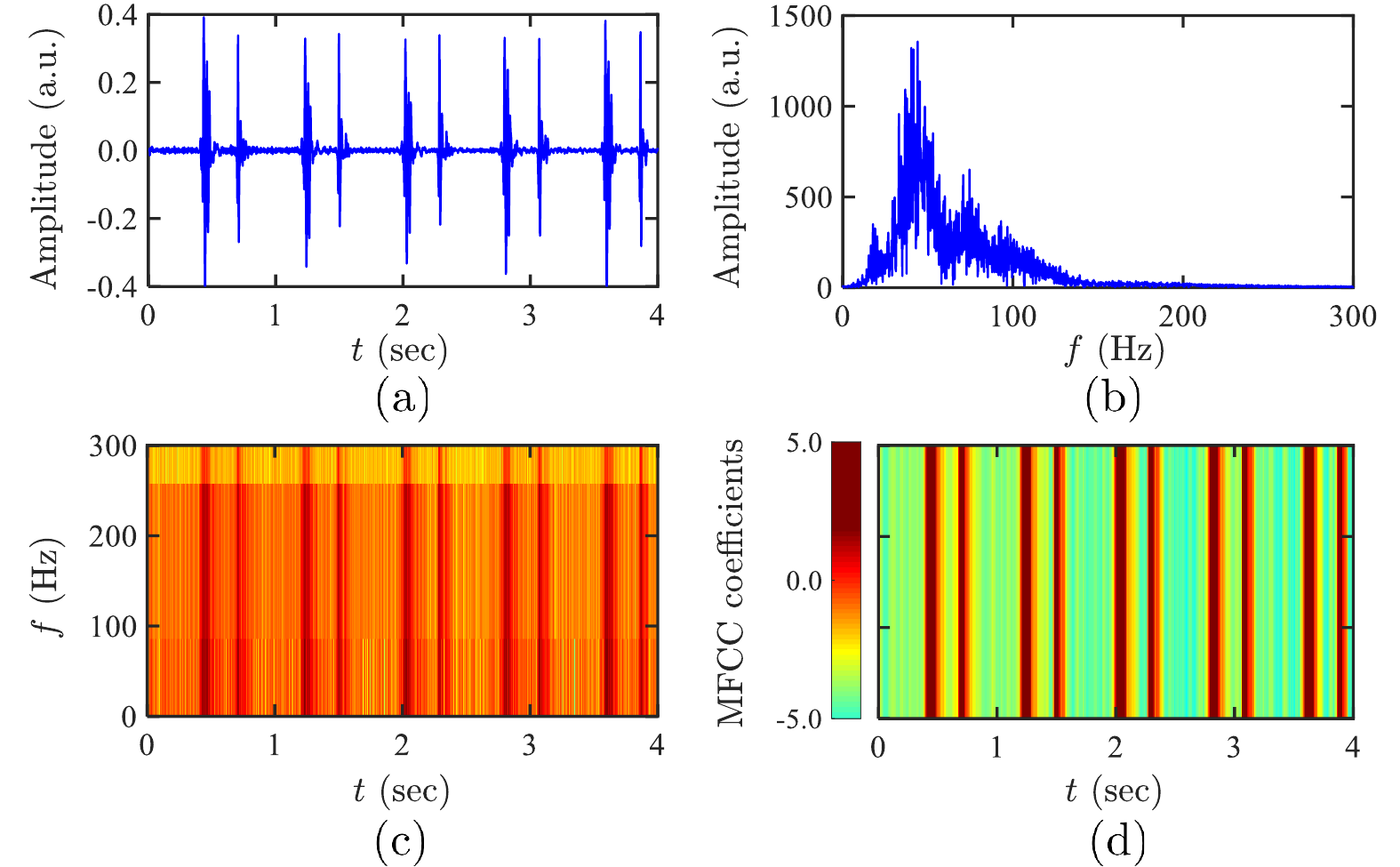}
    \caption{(a) Graphical visualization of raw phonocardiogram (PCG) signal collected using the proposed PhonoTrack device. (b) Fast Fourier Transform (FFT) spectrum, (c) Spectrogram analysis, and (d) Mel-Frequency Cepstral Coefficients (MFCCs) of the corresponding raw PCG signal. For all of them, a four-second time frame was considered.}
    \label{Fig4M}
\end{figure}

The key features for validating the PCG database include the frequency range and distinguishable peaks in heart sounds \cite{liu2016open}. To assess the frequency range for each subject within our proposed PCG dataset, we performed an FFT technique, which converts the time-domain signals into frequency-domain representations. Although graphical representations of raw PCG signals offer initial insights into heart sounds, incorporating time-frequency analysis and continuous wavelet transforms significantly enhances the ability to precisely identify the positions of heart sound peaks concerning both time and frequency. Figure \ref{Fig4M}(a) displays the raw PCG signal collected utilizing the proposed PhonoTrack device. This signal exhibited explicit heart sound peaks with very low baseline noise, indicating an improved signal collection system. Based on Fig.~\ref{Fig4M}(b), we estimated the corresponding frequency range to be approximately 11 to 250 Hz. The existing literature has reported a frequency range of 20 to 250 Hz, which aligns with our findings, thereby supporting the validity and reliability of the collected PCG signal \cite{ren2024comprehensive,arslan2022automated,hazeri2021classification}. In Fig.~\ref{Fig4M}(b), a small amount of noise was observed in the FFT analysis at low frequencies, near zero. Conversely, there was negligible noise at higher frequencies, leading to an excellent PCG signal free from external interference. 

%
%
%

The spectrogram analysis provided a comprehensive insight into the dynamic changes in the frequency content of the S1 and S2 peaks over time with color-indicating strength, as illustrated in Fig.~\ref{Fig4M}(c). In Spectrogram analysis, the short-time Fourier transform (STFT) for each PCG frame was computed using a window length of 256 samples with 50\% overlap, applying the periodic Hann window function to this specified window length. This spectrogram analysis, utilizing a 256-sample window ($\sim 5.8$ ms at a sampling rate of 44.1 kHz), provides very high temporal resolution, enabling precise identification of rapid events, such as the onsets of heart sounds S1 and S2. However, this approach results in relatively low frequency resolution, which limits the ability to distinguish fine spectral details. Despite this limitation, the technique still captures the broader low-frequency energy of PCG data, which typically ranges from 20 to 200 Hz. This trade-off between time and frequency resolution is a key characteristic of the STFT \cite{pachori2023time}. Detailed considerations of the windowing strategy and temporal–frequency trade-offs in PCG analysis are explained in Supplementary Note 1.

To further enhance the power spectral visualization, we calculated MFCCs by mapping frequencies onto the Mel scale and taking their logarithms, as illustrated in Fig.~\ref{Fig4M}(d). Additionally, Fig.~S3 illustrates the subject-wise MFCC plot for STFT analysis across the dataset. This MFCC analysis emphasized the peak intensity of both S1 and S2, highlighting their distinct presence within the signal. Moreover, the narrow yellowish color strips representing noise, found between the power spectral widths of S1 and S2, were negligible. This observation proves the prominence of S1 and S2 peaks at their respective times and frequency positions concerning the raw PCG signal. Additionally, the distance between the S1 and S2 peaks, corresponding to the systolic duration, was shorter than the distance between the S2 and S1 peaks, referring to the diastolic duration. This phenomenon aligns well with the existing literature, further confirming the reliability of our proposed PCG signal dataset \cite{chowdhury2020time,kui2021heart,ismail2023pcg}.  

%
\begin{figure}[hbt]
    \centering
    \includegraphics[width =\linewidth, center]{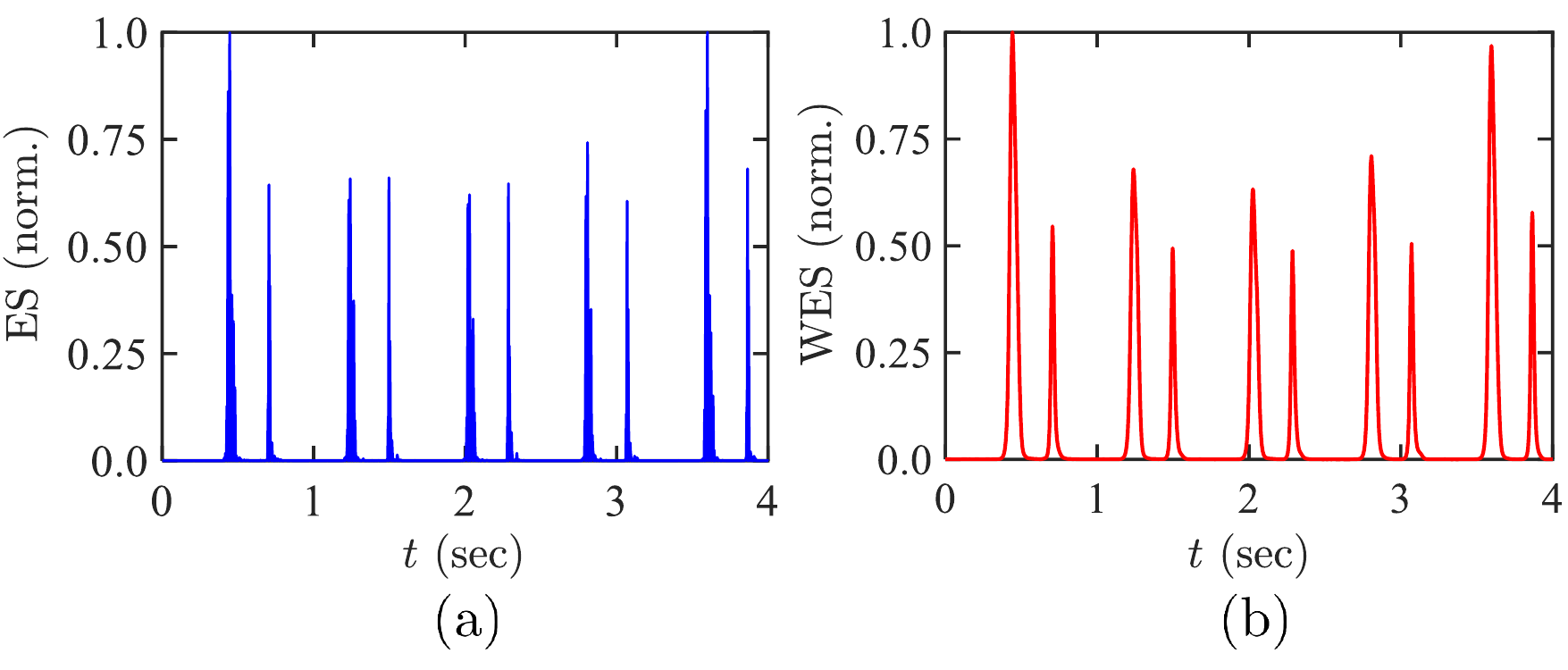}
    \caption{(a) Normalized energy spectrum (ES) and (b) continuous wavelet energy spectrum (WES) based on the same phonocardiogram (PCG) signal as illustrated in Fig.~\ref{Fig4M}(a). For all of them, a four-second time frame was considered.}
    \label{Fig5M}
\end{figure}
%
%

Figure \ref{Fig5M}(a) presents the normalized energy spectrum (ES) based on the same PCG signal shown in Fig.~\ref{Fig4M}(a), where the ES is computed by squaring the amplitude of the raw PCG signal. While this ES analysis highlights the S1 and S2 peaks at the corresponding time frames of the raw PCG signal, the discrete energy spectral widths and ubiquitous amplitude intensities of S1 and S2 peaks can obscure a comprehensive understanding of their durations, rise and decay times, and the sequential amplitude intensity patterns between S1 and S2. To address these challenges, we conducted continuous WES analysis, as depicted in Fig.~\ref{Fig5M}(b). This WES analysis produced a smooth envelope for the S1 and S2 peaks, revealing a consistent pattern wherein the amplitude of S1 is notably greater than S2 across successive heart sound peaks, which aligns well with previously published literature \cite{ergen2012time,meziani2012analysis}. We also calculated data quality metrics through NRMSEs between the WES and ES using three distinct wavelet methods as follows: Morlet, Morse, and Bump. For a PCG signal \( x[n] \), the ES and NRMSE are obtained using the following equations:
\begin{subequations}
\begin{equation}
\label{eq:es}
    \text{ES}(n) = \left| x(n) \right|^2,
\end{equation}
\begin{equation}
\label{eq:nrmse}
    \text{NRMSE} = \sqrt{\frac{\sum_{n=1}^{M}\left| \text{WES}(n)-\text{ES}(n) \right|^2}{\sum_{n=1}^{N}\left| \text{ES}(n) \right|^2}}.
\end{equation}
\end{subequations}

Table \ref{Table2M} presents the corresponding NRMSE values for all subjects. The Morlet wavelet yielded the lowest NRMSE values for each subject, while the Bump wavelet exhibited the highest values, indicating the efficacy of the Morlet wavelet in continuous wavelet analysis. Notably, the NRMSE values for all methods were below one, which was consistent with existing literature \cite{ergen2012time,chowdhury2020time}. The NRMSE values, which are close to unity, indicate that the magnitude difference between the WES and the ES is comparable to the dynamic range of the ES. This phenomenon suggests that the underlying spectral structure remains unchanged, while there may be limited agreement in amplitude. Notably, high NRMSE values can coexist with strong correlation and coherence between the ES and WES. Since these two measures are not mainly affected by linear scaling, they help preserve the relative spectral patterns and frequency-specific organization. Therefore, assessing NRMSE alongside the correlation and coherence between the ES and WES provides a more comprehensive evaluation of the data quality, confirming that structural agreement and amplitude accuracy are regarded as distinct aspects of quality. Additionally, we calculated the Signal-to-Noise Ratio (SNR) to evaluate the quality of the PCG data, as presented in Table \ref{Table2M}. The SNR measures the strength of heart sound signals compared to background noise. Higher SNR values, particularly the average SNR ranging from 22 to 31 dB across subjects, demonstrate that the PCG signals are generally strong with minimal interference from noise, ensuring reliable data quality. These SNR levels are comparable to those reported in previous PCG studies, where values above 20 dB are typically considered sufficient for accurate heart sound analysis \cite{giordano2021automated}. Therefore, the observed SNR range confirms that the recordings from all subjects meet acceptable quality standards for effective feature extraction and subsequent analysis. Furthermore, Table \ref{Table2M} illustrates the frequency range for all subjects, which affirms the acceptable frequency range of the PCG signal database. 
%
%
\begin{table}[hbt]
\centering
\caption{The frequency range and normalized root mean square errors (NRMSEs) between the wavelet energy spectrum (WES) and energy spectrum (ES) for all subjects within the proposed phonocardiogram (PCG) dataset.}
\resizebox{0.85\textwidth}{!}{%
\begin{tabular}{ cccccccc }
\Xhline{3\arrayrulewidth}
     Name & Frequency Range & \multicolumn{3}{c}{NRMSEs} & \multicolumn{3}{c}{SNR} \\
     \cline{3-8} 
       & (Hz) & Morlet & Morse & Bump & Max. & Min. & Avg.\\
    \Xhline{2\arrayrulewidth}
   Subject 01 & 12 -- 162 & 0.746 & 0.788 & 0.923 & 33.442 & 29.669 & 31.613 \\
   Subject 02 & 8 -- 267 & 0.867 & 0.906 & 0.972 & 25.359 & 21.939 & 23.883\\
   Subject 03 & 8 -- 216 & 0.749 & 0.794 & 0.914 & 30.295 & 26.531 & 27.913\\
   Subject 04 & 11 -- 243 & 0.832 & 0.878 & 0.960 & 31.131 & 28.071 & 29.117\\
   Subject 05 & 11 -- 254 & 0.758 & 0.826 & 0.949 & 29.235 & 25.784 & 27.606\\  
   Subject 06 &  8 -- 221  & 0.842 & 0.888 & 0.967 & 32.025 & 29.210 & 30.893\\
   Subject 07 & 7 -- 256 & 0.808 & 0.861 & 0.958 & 27.592 & 21.302 & 24.775\\
   Subject 08 & 7 -- 249 & 0.835 & 0.865 & 0.948 & 22.265 & 15.633 & 18.904 \\
   Subject 09 & 15 -- 239 & 0.772 & 0.811 & 0.934 & 31.131 & 26.157 & 30.189\\
   Subject 10 & 13 -- 234 & 0.777 & 0.818 & 0.935 & 34.499 & 28.793 & 32.084\\ 
   Subject 11 & 9 -- 257 & 0.733 & 0.756 & 0.894 & 31.918 & 24.900 & 28.029\\
   Subject 12 & 6 -- 207 & 0.789 & 0.825 & 0.930 & 24.782 & 18.641 & 21.475\\
   Subject 13 & 12 -- 224 & 0.854 & 0.885 & 0.962 & 27.055 & 19.507 & 23.110\\
   Subject 14 & 10 -- 215 & 0.852 & 0.900 & 0.971 & 24.782 & 18.000 & 22.328 \\
   Subject 15 & 8 -- 234 & 0.834 & 0.878 & 0.955 & 25.632 & 19.432 & 22.412\\   
    \Xhline{3\arrayrulewidth}
\end{tabular}}
\label{Table2M}
\end{table}
%
%

The dataset developed in this study had several limitations. Firstly, it comprised data from only 15 male subjects, which limits the statistical strength and generalizability of the findings. It is important to note that the cohort of solely healthy male adults and the small sample size may limit the applicability of the results to broader populations, particularly when investigating BP estimation. Secondly, the data were collected by the PhonoTrack system under controlled and uniform recording conditions, without any physiological disturbances. Additionally, the relatively high sampling frequency increases data storage requirements. Although the subjects' ages ranged from 18 to 50 years, introducing some variation in body weight and BMI, the diversity remains relatively low. Therefore, while the dataset offers valuable insights under controlled conditions, caution should be maintained when applying these findings to more diverse and real-world settings. The dataset is mainly suitable for evaluating the proposed PhonoTrack system and its methodology for measuring HR and BP under optimal signal conditions. While similar sample sizes and recording durations are considered acceptable for system-level validation as reported in previous literature, broader validation will necessitate larger and more diverse cohorts for application in clinical settings \cite{movahedi2023hardware,kazemnejad2024open}.

%

%
%
\section{Results and discussion}
In this section, we discuss the results obtained using the proposed signal processing-based algorithm to determine HR and BP. The results are described in detail in the following subsection.
%
\begin{figure}[hbt]
    \centering
    \includegraphics[width =\linewidth, center]{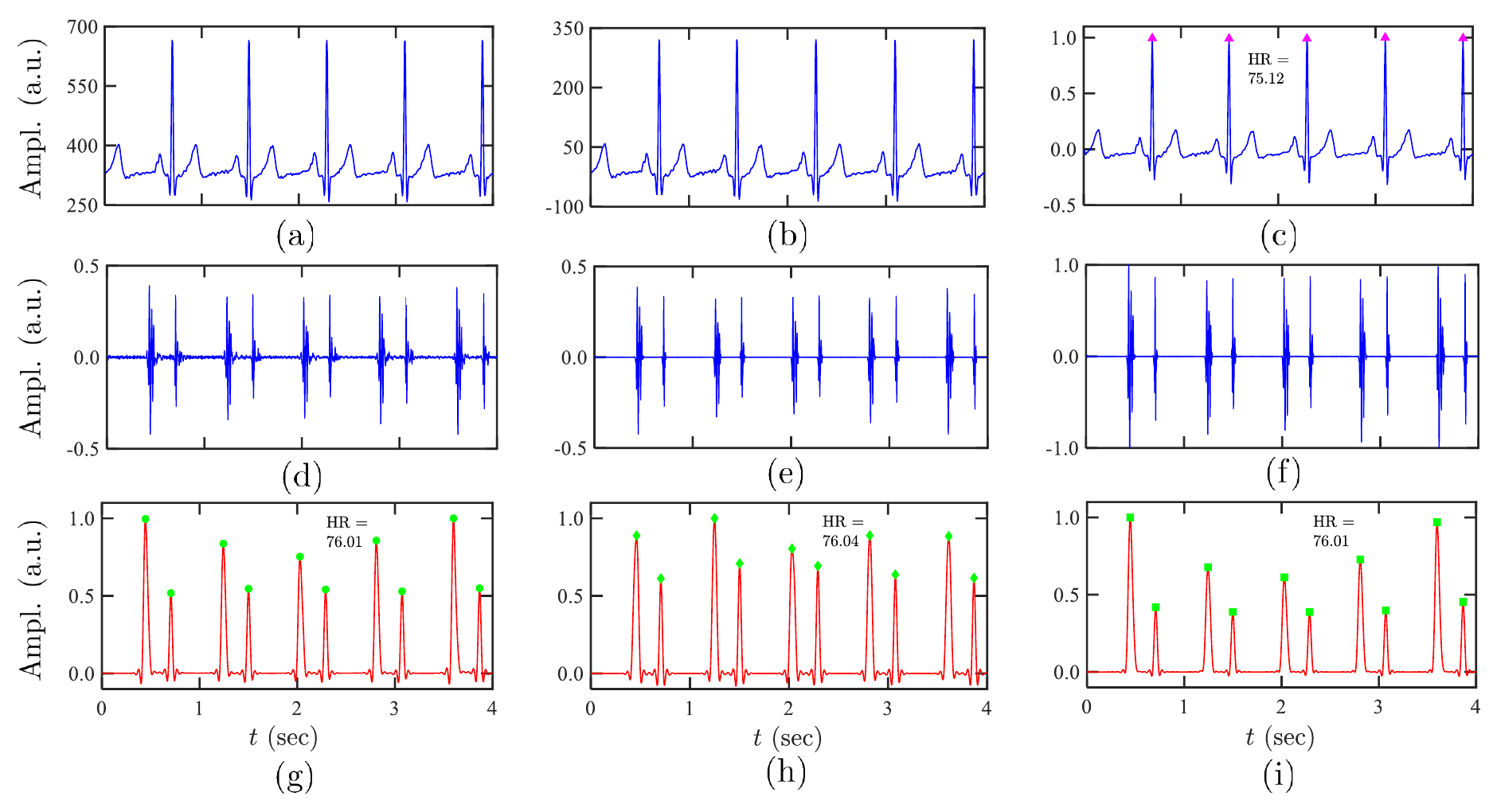}
    \caption{Heart rate (HR) estimation from electrocardiogram (ECG) and phonocardiogram (PCG) signals: (a) Raw ECG, (b) Detrended ECG, (c) Peak detection and heart rate estimation, (d) Raw PCG, (e) Denoised PCG, (f) Normalized PCG, (g) Detected envelope using Hilbert transform (HT), (h) Detected envelope using Shannon Energy (SE) function, and (i) Detected envelop using wavelet energy spectrum (WES). For all of them, a four-second time frame was considered.}
    \label{Fig6M}
\end{figure}
%

%
\subsection{Heart Rate Estimation Performance}

The step-by-step process of the HR estimation methodology is visually illustrated in Fig.~\ref{Fig6M}. Specifically, Fig.~\ref{Fig6M}(a) presents the raw ECG signal, while Fig.~\ref{Fig6M}(b) shows the detrended version of the ECG, which helps eliminate baseline wander and low-frequency noise. Subsequently, Fig.~\ref{Fig6M}(c) illustrates the peak detection stage, where R-peaks are identified to calculate the HR. This HR derived from the ECG signal was treated as the ground truth for validating the HR estimated from the corresponding PCG signal over the same time frame. The next part of the process focused on PCG signal preparation. Figs.~\ref{Fig6M}(d)--(f) depict the raw PCG signal, the denoised version after wavelet-based denoising, and the final normalized PCG signal, respectively. As previously mentioned, to extract HR from the PCG, we computed the signal envelope using three distinct methods. The resulting envelopes, along with their associated HR estimations, are shown in Figs.~\ref{Fig6M}(g)--(i).

\begin{figure}[hbt]
    \centering
    \includegraphics[width =1.0\linewidth, center]{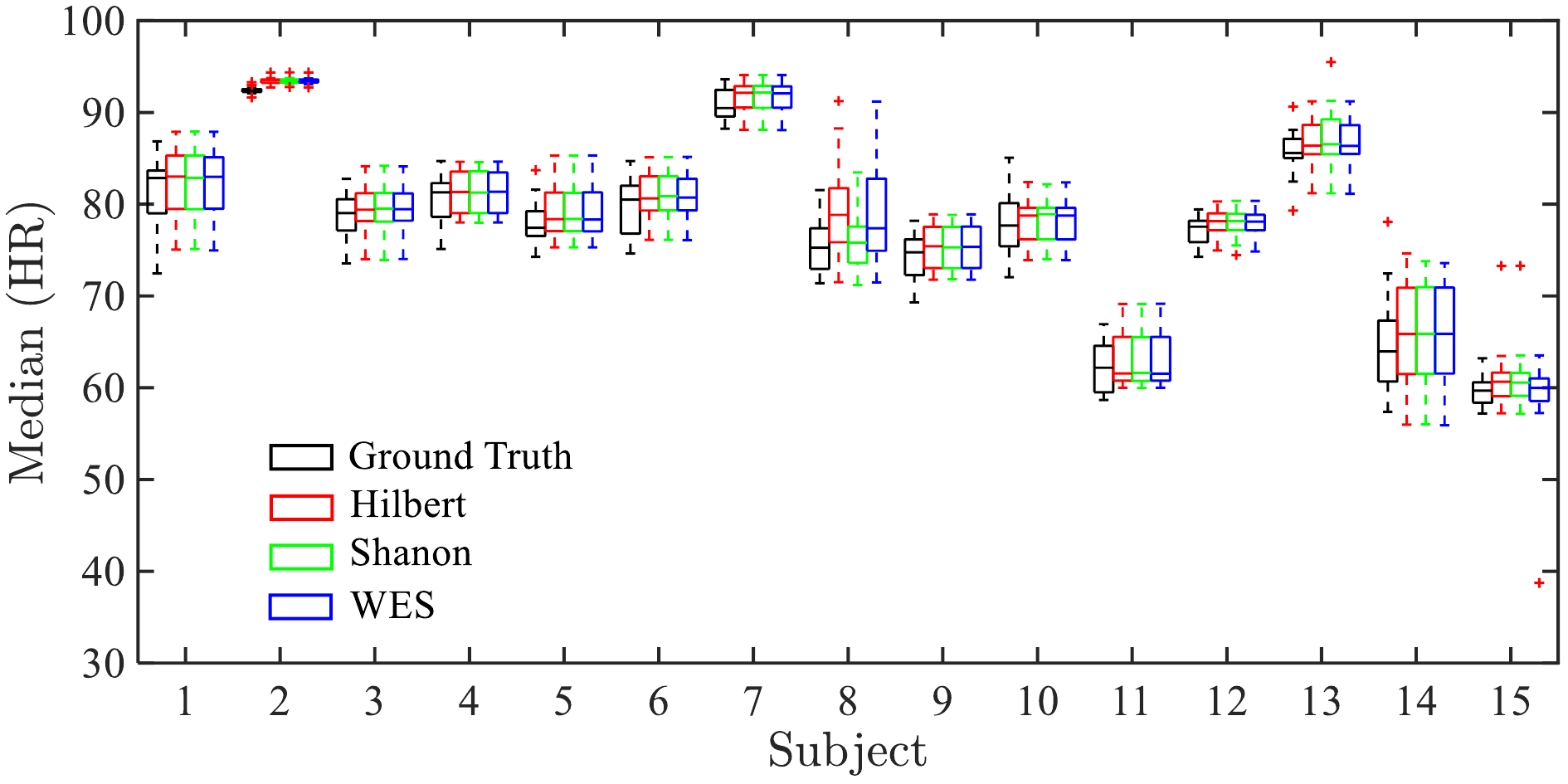}
    \caption{Box plot for all subjects using three envelope detection methods in heart rate (HR) estimation.}
    \label{Fig7M}
\end{figure}

To comprehensively evaluate the performance of our proposed algorithm for estimating HR from PCG signals, we employed both a box plot and a Pearson's correlation coefficient plot. These visualizations provided valuable insights into the accuracy, consistency, and reliability of the HR estimation in comparison to the ground truth, which was derived from the ECG signals. The box plot, as illustrated in Fig.~\ref{Fig7M}, compares the HR values predicted by the three different envelope-based methods against the ground truth across all subjects. This graphical representation highlights essential statistical metrics, including the median, inter-quartile range (IQR), and the presence of potential outliers. A narrow IQR with fewer or no outliers typically indicates a more robust and consistent estimation method, whereas a wider IQR or the presence of extreme values may point to greater variability or noise sensitivity. In our analysis, Subject 2 exhibited the smallest IQR, demonstrating high consistency and accuracy in HR prediction across all three methods. Conversely, Subjects 8 and 14 showed the largest IQRs, suggesting higher variability, possibly due to signal artifacts or subject-specific physiological factors. To investigate the cause, we inspected the frame-wise signal quality measures produced by our pipeline (SNR computed per frame) and found that these subjects contained some low-quality frames (lower mean SNR) compared to the low-IQR subjects. For Subjects 8 and 14, most frames produced reasonable HR estimates, but a few frames had unusually large errors. This suggests the variability comes from short periods of noise or temporary sensor/contact disturbances, rather than consistent failure throughout the recording. For the remaining subjects, the IQRs were relatively uniform and fell within an acceptable range, indicating the general robustness of our methods across the study subjects. Furthermore, it is noteworthy that for all subjects, the median HR values obtained from the proposed methods closely align with the ground truth median values derived from ECG data. This strong agreement, clearly observed in Fig.~\ref{Fig7M}, highlights the effectiveness and reliability of our PCG-based HR estimation methods in approximating the actual HR with high precision.

%
%

\begin{figure}[hbt]
    \centering
    \includegraphics[width =0.81\linewidth, center]{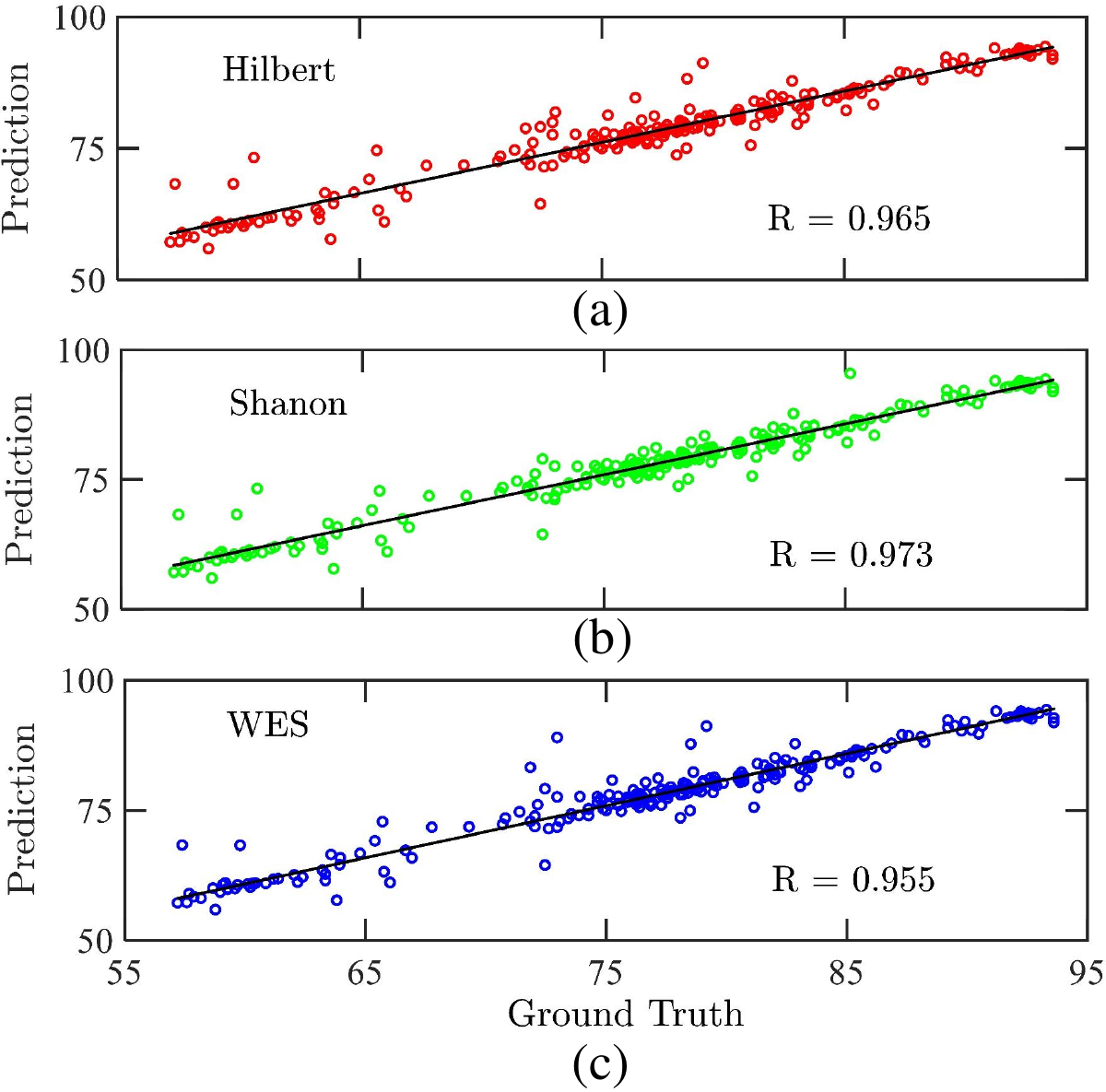}
    \caption{Heart rate estimation performance with Pearson's correlation plot for three envelope detection methods as follows: (a) Hilbert Transform; (b) Shannon Energy Function; and (c) Wavelet Energy Spectrum.}
    \label{Fig8M}
\end{figure}

To further demonstrate the effectiveness of the proposed method, we included Pearson's correlation plot in Fig.~\ref{Fig8M}, which illustrates the relationship between the estimated HR and the ground truth values. Pearson's correlation coefficient ($R$) quantifies the strength and direction of the linear relationship between the two variables, with $R = 1$ indicating a perfect positive correlation, R = 0 suggesting no correlation, and $R = -1$ representing a perfect negative correlation. A high positive value of $R$ close to 1 in Fig.~\ref{Fig8M} confirms that the estimated HR closely aligns with the actual HR, demonstrating the accuracy and reliability of the proposed method. For the three envelope detection methods in HR estimation, we calculated Pearson's correlation coefficients of 0.965, 0.973, and 0.955 for the HT, SE, and WES-based methods, respectively.

In addition to correlation, we report frame-wise error metrics and agreement analysis using ECG-derived HR as the reference. Each 60-s ECG/PCG recording is segmented into non-overlapping 4-s frames, yielding 15 frames per subject (15 subjects; 225 frames in total). For each frame, HR is estimated from the PCG using the three envelope pipelines (Hilbert, Shannon energy, and WES), and the frame-wise difference is defined as $\Delta \mathrm{HR}=\mathrm{HR}_{\mathrm{pcg}}-\mathrm{HR}_{\mathrm{ecg}}$. To keep the evaluation subject-wise and avoid overweighting any individual recording, we first compute MAE, RMSE, and the mean $\pm$ SD of $\Delta \mathrm{HR}$ by aggregating the 15 frames within each subject, and then report the average of these subject-level metrics across all 15 subjects. Across subjects, the Hilbert-envelope method achieves MAE $=1.824$ bpm and RMSE $=2.467$ bpm with a mean difference of $0.962 \pm 1.939$ bpm, the Shannon-energy method achieves MAE $=1.330$ bpm and RMSE $=1.688$ bpm with a mean difference of $0.880 \pm 1.304$ bpm, and the WES method achieves MAE $=1.517$ bpm and RMSE $=1.992$ bpm with a mean difference of $1.069 \pm 1.549$ bpm. Subject-wise results are provided in Supplementary Table~S1.

In addition, we used Bland-Altman analysis, as shown in Figure S4 of the supplementary materials, to assess agreement between the estimated HR and the ECG reference by visualizing $\Delta \text{HR}$ against the mean HR and reporting the bias and 95\% limits of agreement. Based on the Bland–Altman plots (Fig.~S4), the HR estimates show low systematic bias and tight agreement with the ECG reference, as most frame-level points lie within the 95\% limits of agreement. A small number of outliers are present, likely corresponding to transient low-quality frames (e.g., brief noise/contact artifacts), but the overall dispersion remains limited across the HR range.
%
Notably, the Shannon-entropy method shows the best overall accuracy in our experiments, achieving the lowest MAE and RMSE and the smallest mean bias compared to the ECG reference. Since the three methods are close in performance, we prioritize the Shannon-energy envelope extraction pipeline for real-time wearable operation as it combines slightly better error statistics with a lightweight time-domain implementation. The Hilbert-envelope method remains a strong alternative with comparable accuracy and similarly low computational cost, whereas the WES-based approach is more computation-intensive and is better suited for offline analysis or optional quality checking.

To probe robustness beyond controlled recordings, we additionally tested the proposed HR pipelines on some external PCG samples from the PhysioNet 2016~\cite{liu2016open} dataset and on a small set of our own recordings captured in a noisier environment. The resulting error statistics are summarized in Supplementary Table~S2, indicating that the envelope-based methods remain reliable under moderate acoustic noise, with occasional outlier frames occurring during transient disturbances. We note that a dedicated ambulatory (walking/running) study is a natural next step for future work.
%
%

\subsection{Blood Pressure Estimation Performance}

To estimate BP solely from the PCG signal, we extracted a set of informative features that captured the underlying characteristics of cardiac activity. These features were then utilized to perform a multiple regression analysis, which resulted in the formulation of a semi-empirical model capable of predicting both SBP and DBP. The regression approach allowed us to identify the most significant contributors to BP estimation based on their statistical relationship with the ground truth measurements. Table \ref{Table1M} presents the coefficients derived from the multiple regression analysis, representing the value of each feature in the final semi-empirical model. These coefficients play a crucial role in quantifying the contribution of individual PCG-derived features toward the estimation of SBP and DBP values.

%
\begin{figure}[hbt]
    \centering
    \includegraphics[width =0.97\linewidth, center]{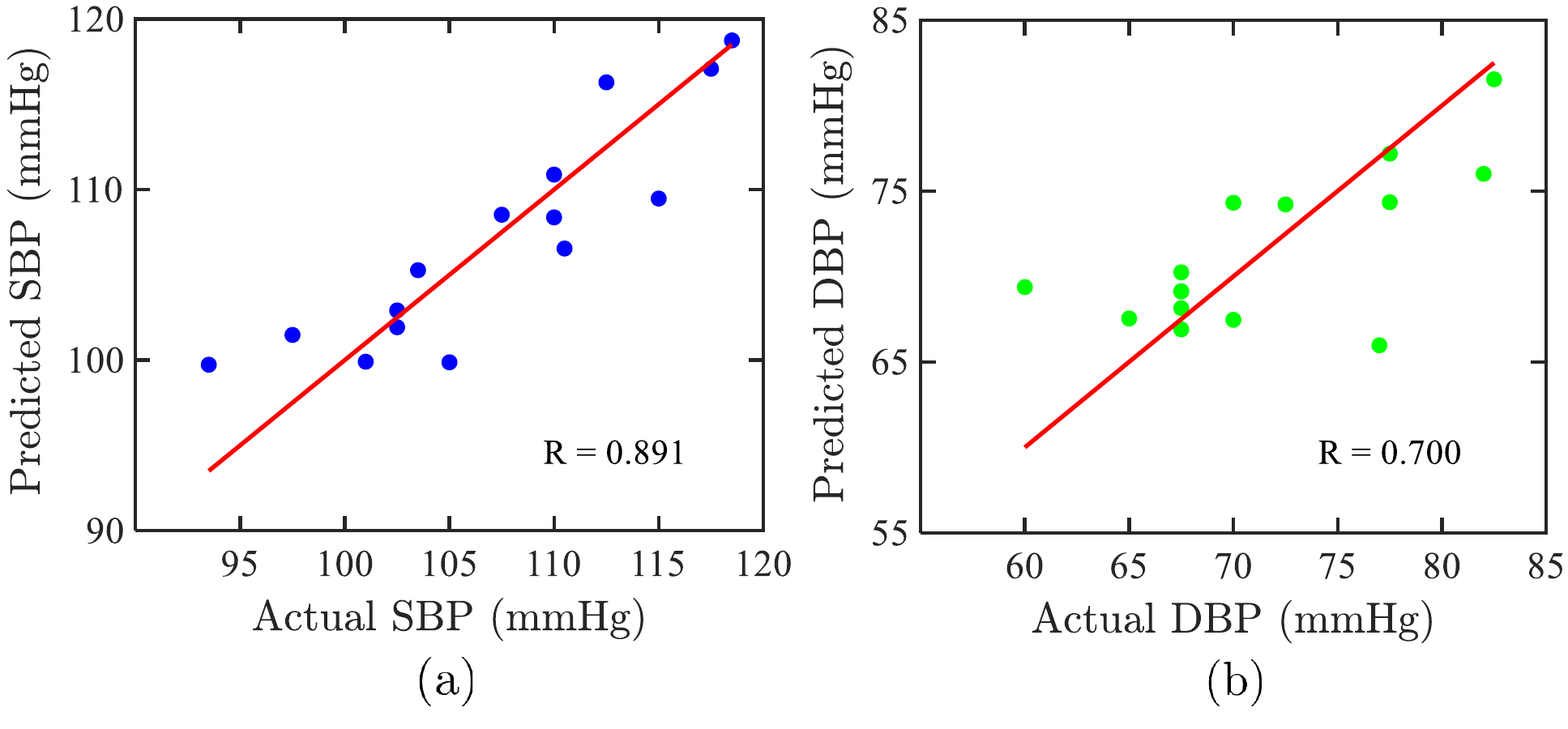}
    \caption{Blood pressure (BP) estimation result, along with Pearson's coefficient using multiple regression for predicting the values of (a) systolic blood pressure (SBP) and (b) diastolic blood pressure (DBP).}
    \label{Fig9M}
\end{figure}
%
To evaluate the performance of the proposed model, Fig.~\ref{Fig9M} illustrates a Pearson's correlation plot comparing the predicted BP values to the ground truth. The plot demonstrates a strong alignment between the estimated and actual values, with most data points closely distributed along the 45-degree reference line. This visual alignment indicates a strong positive correlation and confirms the effectiveness of the model in capturing the relationship between the PCG features and true BP measurements. Quantitatively, the model achieved a Pearson's correlation coefficient of $R = 0.891$ for SBP and $R = 0.700$ for DBP. In the same evaluation setting, we also report the dispersion of the absolute BP error in mmHg, achieving an error standard deviation of 2.10~mmHg and 3.20~mmHg for SBP and DBP across all subjects, respectively. These values reflect a high level of accuracy in SBP prediction and a moderate but useful correlation in DBP estimation, validating the potential of our semi-empirical approach in non-invasive and continuous BP monitoring applications based solely on the PCG signal.

To evaluate whether the BP regression generalizes across individuals, we use LOOCV at the subject level. Since each subject contributes one averaged SBP/DBP value, we treat subjects as the independent samples. For each fold, one subject is excluded from training, the regression model is fitted using the remaining subjects, and SBP/DBP are predicted for the held-out subject. This is repeated for all subjects, and performance is computed using only the held-out predictions. Using subject-wise LOOCV, the proposed empirical model yields RMSE = 9.8871 mmHg and MAE = 7.2565 mmHg for SBP, and RMSE = 11.0508 mmHg and MAE = 8.6295 mmHg for DBP.

%
\begin{figure}[hbt]
    \centering
    \includegraphics[width =0.95\linewidth, center]{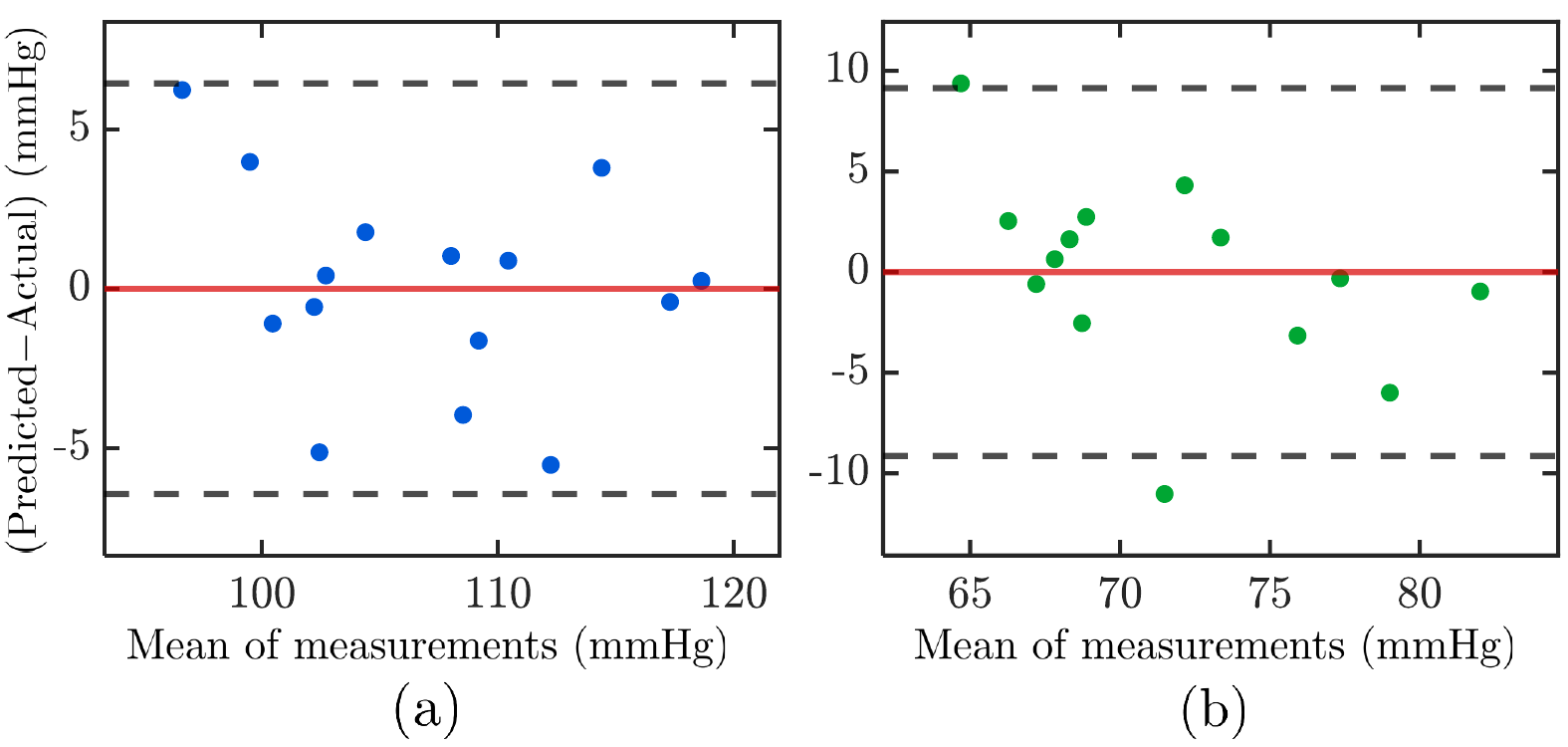}
    \caption{Bland--Altman plots for BP estimation. (a) SBP and (b) DBP agreement between estimated and reference measurements; red line indicates mean bias and dashed lines show the 95\% limits of agreement (bias $\pm 1.96\,\mathrm{SD}$).}
    \label{Fig10M}
\end{figure}
%

To further examine agreement and identify potential bias patterns, we added Bland-Altman plots for both SBP and DBP (Fig.~\ref{Fig10M}). In these plots, the vertical axis represents the difference between predicted and reference BP in mmHg, and the horizontal axis shows the mean of the two measurements. The red horizontal line represents the mean error (bias), and the dashed lines indicate the 95\% limits of agreement. For SBP (Fig.~\ref{Fig10M}(a)), the differences are centered close to zero, indicating minimal systematic overestimation or underestimation, and most points lie within the limits of agreement, suggesting consistent estimation across the observed SBP range. For DBP (Fig.~\ref{Fig10M}(b)), the spread of errors is wider and includes a few larger negative deviations, which explains the higher RMSE/MAE for DBP. Importantly, the Bland-Altman plots complement the correlation results by showing that the model's errors are largely bounded and centered around zero, while also highlighting that DBP estimation remains more variable than SBP.

%
%
\begin{table}[H]
\centering
\caption{Comparison of the performance of the proposed PhonoTrack system in estimating heart rate (HR) and blood pressure (BP) with existing literature.}
\resizebox{1.0\textwidth}{!}{%
\begin{tabular}{c c c c c c c}
\Xhline{3\arrayrulewidth}
    System Name & Sample Size & HR Estimation Accuracy & BP Estimation Error & Ref. \\
    & & (Pearson coefficient, R) & (Standard Deviation, SD (mmHg)) & \\
    \Xhline{2\arrayrulewidth}
    In-ear PPG Sensor    & $97$ & $0.83$ & Not reported & \cite{adams2022accurate} \\
    PPG wearable sensors & $31$ & $0.834$ & Not reported & \cite{coste2025comparative} \\
    Fetal PCG HR estimation & Not specified & $0.90$ & Not reported & \cite{dia2019fetal} \\
    EPHNOGRAM system & $68$ & $\sim 0.65$ & Not reported & \cite{kazemnejad2024open} \\
    Contact-type HR sensor & $40$ & $\sim 0.95$ & Not reported & \cite{zang2025novel} \\     
    Wearable cuff‑less watch    & $21$ & Not reported & SD$_{\rm SBP}=4.70$ \& SD$_{\rm DBP}=2.70$ & \cite{ganti2020wearable} \\
    PCG$+$ANN BP Estimation    & $37$ & Not reported & SD$_{\rm SBP}=13.31$ \& SD$_{\rm DBP}=9.52$ & \cite{omari2019new} \\
    BP monitoring system    & $12$ & Not reported & SD$_{\rm SBP}=3.139$ \& SD$_{\rm DBP}=5.198$ & \cite{chen2017calculating} \\   
    PhonoTrack System    & $15$ & $0.973$ & SD$_{\rm SBP}=2.10$ \& SD$_{\rm DBP}=3.20$ & This work \\
    \Xhline{3\arrayrulewidth}
\end{tabular}}
\label{Table3M}
\end{table}
%
%

A comprehensive comparison with existing literature highlights the significance of the proposed PhonoTrack device. Previous studies have shown that the synchronized acquisition of PCG and ECG signals is both feasible and promising for cardiac monitoring. For example, Damani et al.~investigated simultaneous ECG and PCG recordings using a digital stethoscope to assess electromechanical cardiac activity \cite{damani2021investigation}. However, their focus was primarily on characterizing the signals rather than on extensive physiological modeling. Similarly, a prototype for a wearable sensor measured heart rate during exercise, achieving an intraclass correlation coefficient (ICC) of $0.96$ at rest and $\sim 0.92$ for the average test heart rate \cite{loro2024validation}. Moreover, an in-ear PPG sensor reported a Pearson correlation of $0.83$ for HR estimation in relation to ECG \cite{adams2022accurate}. Conversely, the PhonoTrack system achieved the Pearson correlations of HR calculations up to $0.973$ using multiple signal-processing approaches, demonstrating strong agreement with ECG-derived HR estimation. Table \ref{Table3M} summarizes the comparison of HR and BP estimation performance of the proposed PhonoTrack system with relevant reported literature.

Most cuffless blood pressure estimation methods rely on pulse transit time and ECG--PPG features, which adds complexity to the system. However, the semi-empirical multiple regression model developed in this study estimated systolic and diastolic blood pressures using only PCG features, achieving competitive accuracy with a simpler and more cost-effective setup. Unlike earlier PCG-based devices that focused on heart sound segmentation and murmur detection, the PhonoTrack device offers superior HR accuracy and excellent BP estimation through lower-complexity approaches, such as straightforward signal processing and spectral and wavelet analyses. This proposed device also provides an integrated, validated, and portable solution for multimodal cardiovascular monitoring. However, the current findings should be interpreted as preliminary proof-of-concept results rather than evidence of clinical readiness. Given the limited cohort, short recording duration, and manual cuff-based blood pressure measurements, the present study is insufficient to support claims related to cardiovascular disease assessment, clinical deployment, or routine clinical use.

The cost-effectiveness of the proposed real-time PhonoTrack system stems from its simple design and the use of low-cost, readily available components. The system features a mechanical diaphragm combined with a basic wired microphone, which is securely attached to the chest using a strap. This design eliminates the need for expensive digital stethoscopes and multi-sensor platforms. An embedded microcontroller manages signal acquisition and processes data in real time. The physiological parameters are displayed through an intuitive monitoring interface, reducing the need for proprietary clinical systems. By minimizing hardware complexity and leveraging software-driven processing on an embedded platform, the PhonoTrack device offers a cost-effective alternative to traditional cardiac monitoring systems, making it suitable for scalable clinical environments. An approximate component-wise cost breakdown of the proposed prototype is provided in Table~S4 of the Supplementary Materials. However, these prices are indicative only and may vary depending on supplier, purchase quantity, and regional availability.

Future studies on the PhonoTrack device will focus on validating its effectiveness across a more diverse range of populations and real-world situations. Participant groups will include individuals of various ages, balanced genders, differing body mass indices, and those with cardiovascular conditions to ensure a thorough assessment \cite{ahmad2025advancements}. Data collection will take place during daily activities, mild exertion, changes in posture, and stressful situations, allowing for the evaluation of the system's performance amid motion artifacts and noisy environments. Additionally, investigations into real-time signal acquisition and on-device processing will facilitate continuous monitoring and immediate feedback. These efforts intend to transition PhonoTrack from a laboratory prototype to a practical solution for early detection, routine screening, and long-term management in clinical settings.

%
%
\section{Conclusion}

As the global prevalence of cardiovascular diseases continues to rise, the demand for accessible, reliable, and continuous cardiac health monitoring solutions becomes increasingly urgent. In response to this growing need, our study proposed a dedicated PCG (phonocardiogram) signal database, acquired using a custom-designed and cost-effective acoustic sensor integrated within the portable PhonoTrack device. This lightweight and user-friendly data acquisition system allows for the collection of high-quality cardiac sounds, facilitating real-time analysis and monitoring. We utilized FFT and MFCC analyses to validate the database. Additionally, we developed simple signal processing algorithms that effectively estimate HR and BP directly from the PCG signal. Our approach demonstrated strong performance, achieving a peak $R$ of 0.973 for HR estimation. For blood pressure prediction, our semi-empirical model showed $R$ of 0.891 for SBP and 0.700 for DBP, indicating good agreement with the ground truth recorded using a sphygmomanometer-based BP measurement system. The promising results of this study highlight the potential of PCG-based monitoring systems as a viable alternative for non-invasive, continuous cardiovascular assessment. Our proposed framework lays a solid foundation for the future development of fully functional, low-cost, and wearable devices that can significantly enhance home-based and outpatient cardiac care. The potential of PhonoTrack in clinical settings is promising, and future research is required to address current limitations by expanding the dataset to include a larger and more diverse population, specifically incorporating female subjects and covering a wider range of ages and BMIs. Furthermore, PCG recordings will be conducted during various physiological states, including exercise, postural changes, and stress. While previous studies have validated similar devices with comparable sample sizes, achieving accurate HR and BP estimations in broader clinical applicability will necessitate considerably larger and more diverse cohorts, underscoring significant directions for future research.

%
%
%

\section*{Data Availability} 
All the simulation data are presented in the main text, and the corresponding author will provide it upon reasonable request. 
%

\section*{Author Declaration} The authors have no conflicts to disclose.

%
%
\section*{Acknowledgment} 
The authors would like to express their sincere gratitude to all the volunteers who participated in this study. Their willingness to contribute their time and physiological data was essential to the success of this research.

%
%
\small
\bibliographystyle{ieeetr}
\bibliography{references}

@article{ahmad2025advancements,
  title={Advancements in wearable heart sounds devices for the monitoring of cardiovascular diseases},
  author={Ahmad, Rafi u Shan and Khan, Muhammad Shehzad and Hilal, Mohamed Elhousseini and Khan, Bangul and Zhang, Yuanting and Khoo, Bee Luan},
  journal={SmartMat},
  volume={6},
  number={1},
  pages={e1311},
  year={2025},
  publisher={Wiley Online Library}
}

@article{mathew2024foundation,
  title={Foundation models for cardiovascular disease detection via biosignals from digital stethoscopes},
  author={Mathew, George and Barbosa, Daniel and Prince, John and Venkatraman, Subramaniam},
  journal={npj Cardiovascular Health},
  volume={1},
  number={1},
  pages={25},
  year={2024},
  publisher={Nature Publishing Group UK London}
}

@article{zang2025novel,
  title={A novel wearable device integrating \textsc{ECG} and \textsc{PCG} for cardiac health monitoring},
  author={Zang, Junbin and An, Qi and Li, Bo and Zhang, Zhidong and Gao, Libo and Xue, Chenyang},
  journal={Microsystems \& Nanoengineering},
  volume={11},
  number={1},
  pages={7},
  year={2025},
  publisher={Nature Publishing Group UK London}
}

@article{gudigar2024automated,
  title={Automated system for the detection of heart anomalies using phonocardiograms: A systematic review (2013-2023)},
  author={Gudigar, Anjan and Raghavendra, U and Maithri, M and Samanth, Jyothi and Inamdar, Mahesh Anil and Vidhya, V and Jahmunah, V and Prabhu, Mukund A and Tan, Ru-San and Yeong, Chai Hong and others},
  journal={IEEE Access},
  year={2024},
  publisher={IEEE}
}

@article{hanna2002history,
  title={A history of cardiac auscultation and some of its contributors},
  author={Hanna, Ibrahim R and Silverman, Mark E},
  journal={The American Journal of Cardiology},
  volume={90},
  number={3},
  pages={259--267},
  year={2002},
  publisher={Elsevier}
}

@article{pelech2004physiology,
  title={The physiology of cardiac auscultation},
  author={Pelech, Andrew N},
  journal={Pediatric Clinics},
  volume={51},
  number={6},
  pages={1515--1535},
  year={2004},
  publisher={Elsevier}
}

@article{ren2018novel,
  title={A novel cardiac auscultation monitoring system based on wireless sensing for healthcare},
  author={Ren, Haoran and Jin, Hailong and Chen, Chen and Ghayvat, Hemant and Chen, Wei},
  journal={IEEE Journal of Translational Engineering in Health and Medicine},
  volume={6},
  pages={1--12},
  year={2018},
  publisher={Ieee}
}

@article{ren2024comprehensive,
  title={A comprehensive survey on heart sound analysis in the deep learning era},
  author={Ren, Zhao and Chang, Yi and Nguyen, Thanh Tam and Tan, Yang and Qian, Kun and Schuller, Bj{\"o}rn W},
  journal={IEEE Computational Intelligence Magazine},
  volume={19},
  number={3},
  pages={42--57},
  year={2024},
  publisher={IEEE}
}

@article{ismail2018localization,
  title={Localization and classification of heart beats in phonocardiography signals--a comprehensive review},
  author={Ismail, Shahid and Siddiqi, Imran and Akram, Usman},
  journal={EURASIP Journal on Advances in Signal Processing},
  volume={2018},
  number={1},
  pages={1--27},
  year={2018},
  publisher={Springer}
}

@article{shao2016simultaneous,
  title={Simultaneous monitoring of ballistocardiogram and photoplethysmogram using a camera},
  author={Shao, Dangdang and Tsow, Francis and Liu, Chenbin and Yang, Yuting and Tao, Nongjian},
  journal={IEEE Transactions on Biomedical Engineering},
  volume={64},
  number={5},
  pages={1003--1010},
  year={2016},
  publisher={IEEE}
}

@article{pinheiro2010survey,
  title={A survey on unobtrusive measurements of the cardiovascular function and their practical implementation in wheelchairs},
  author={Pinheiro, Eduardo and Postolache, Octavian and Gir{\~a}o, Pedro},
  journal={Sensors \& Transducers},
  volume={9},
  pages={182},
  year={2010},
  publisher={IFSA Publishing, SL}
}

@article{hussain2022modern,
  title={Modern diagnostic imaging technique applications and risk factors in the medical field: a review},
  author={Hussain, Shah and Mubeen, Iqra and Ullah, Niamat and Shah, Syed Shahab Ud Din and Khan, Bakhtawar Abduljalil and Zahoor, Muhammad and Ullah, Riaz and Khan, Farhat Ali and Sultan, Mujeeb A},
  journal={BioMed Research International},
  volume={2022},
  number={1},
  pages={5164970},
  year={2022},
  publisher={Wiley Online Library}
}

@article{monteiro2023novel,
  title={A novel approach to simultaneous phonocardiography and electrocardiography during auscultation},
  author={Monteiro, Sofia M and Da Silva, Hugo Pl{\'a}cido},
  journal={IEEE Access},
  volume={11},
  pages={78224--78236},
  year={2023},
  publisher={IEEE}
}

@article{ming2020continuous,
  title={Continuous physiological monitoring using wearable technology to inform individual management of infectious diseases, public health and outbreak responses},
  author={Ming, Damien K and Sangkaew, Sorawat and Chanh, Ho Q and Nhat, Phung TH and Yacoub, Sophie and Georgiou, Pantelis and Holmes, Alison H},
  journal={International Journal of Infectious Diseases},
  volume={96},
  pages={648--654},
  year={2020},
  publisher={Elsevier}
}

@article{gambhir2021continuous,
  title={Continuous health monitoring: An opportunity for precision health},
  author={Gambhir, Sanjiv S and Ge, T Jessie and Vermesh, Ophir and Spitler, Ryan and Gold, Garry E},
  journal={Science Translational Medicine},
  volume={13},
  number={597},
  pages={eabe5383},
  year={2021},
  publisher={American Association for the Advancement of Science}
}

@inproceedings{podaru2020blood,
  title={Blood pressure estimation based on synchronous ECG and PPG recording},
  author={Podaru, Alexandru Constantin and David, Valeriu},
  booktitle={2020 International Conference and Exposition on Electrical And Power Engineering (EPE)},
  pages={640--645},
  year={2020},
  organization={IEEE}
}

@article{gonzalez2022estimation,
  title={Estimation of systolic blood pressure by Random Forest using heart sounds and a ballistocardiogram},
  author={Gonzalez-Landaeta, Rafael and Ramirez, Brenda and Mejia, Jose},
  journal={Scientific Reports},
  volume={12},
  number={1},
  pages={17196},
  year={2022},
  publisher={Nature Publishing Group UK London}
}

@misc{pascal-chsc-2011,
       author = "Bentley, P. and Nordehn, G. and Coimbra, M. and Mannor, S.",
       title = "The {PASCAL} {C}lassifying {H}eart {S}ounds {C}hallenge 2011 {(CHSC2011)} {R}esults",
       howpublished = "http://www.peterjbentley.com/heartchallenge/index.html"}

@inproceedings{clifford2016classification,
  title={Classification of normal/abnormal heart sound recordings: The PhysioNet/Computing in Cardiology Challenge 2016},
  author={Clifford, Gari D and Liu, Chengyu and Moody, Benjamin and Springer, David and Silva, Ikaro and Li, Qiao and Mark, Roger G},
  booktitle={2016 Computing in Cardiology Conference (CinC)},
  pages={609--612},
  year={2016},
  organization={IEEE}
}

@article{kazemnejad2024open,
  title={An open-access simultaneous electrocardiogram and phonocardiogram database},
  author={Kazemnejad, Arsalan and Karimi, Sajjad and Gordany, Peiman and Clifford, Gari D and Sameni, Reza},
  journal={Physiological Measurement},
  volume={45},
  number={5},
  pages={055005},
  year={2024},
  publisher={IOP Publishing}
}

@article{movahedi2023hardware,
  title={A hardware-software system for accurate segmentation of phonocardiogram signal},
  author={Movahedi, Mohammad Mehdi and Shakerpour, Mohamadreza and Mousavi, Shahrokh and Nori, Ahmad and Dehkordi, Seyyed Hesam Mousavian and Parsaei, Hossein},
  journal={Journal of Biomedical Physics \& Engineering},
  volume={13},
  number={3},
  pages={261},
  year={2023}
}

@article{touahria2023feature,
  title={Feature selection algorithms highlight the importance of the systolic segment for normal/murmur PCG beat classification},
  author={Touahria, Rima and Hacine-Gharbi, Abdenour and Ravier, Philippe},
  journal={Biomedical Signal Processing and Control},
  volume={86},
  pages={105288},
  year={2023},
  publisher={Elsevier}
}

@article{hamza2024comprehensive,
  title={A Comprehensive Overview of Heart Sound Analysis Using Machine Learning Methods},
  author={Hamza, Motaz Faroq A Ben and Sjarif, Nilam Nur Amir},
  journal={IEEE Access},
  year={2024},
  publisher={IEEE}
}

@article{sa2012low,
  title={A low-cost, portable, high-throughput wireless sensor system for phonocardiography applications},
  author={Sa-Ngasoongsong, Akkarapol and Kunthong, Jakkrit and Sarangan, Venkatesh and Cai, Xinwei and Bukkapatnam, Satish TS},
  journal={Sensors},
  volume={12},
  number={8},
  pages={10851--10870},
  year={2012},
  publisher={Molecular Diversity Preservation International (MDPI)}
}

@article{chen2017calculating,
  title={Calculating blood pressure based on measured heart sounds},
  author={Chen, Lingguang and Wu, Sean F and Xu, Yong and Lyman, William D and Kapur, Gaurav},
  journal={Journal of Computational Acoustics},
  volume={25},
  number={03},
  pages={1750014},
  year={2017},
  publisher={World Scientific}
}

@article{liu2016open,
  title={An open access database for the evaluation of heart sound algorithms},
  author={Liu, Chengyu and Springer, David and Li, Qiao and Moody, Benjamin and Juan, Ricardo Abad and Chorro, Francisco J and Castells, Francisco and Roig, Jos{\'e} Millet and Silva, Ikaro and Johnson, Alistair EW and others},
  journal={Physiological Measurement},
  volume={37},
  number={12},
  pages={2181},
  year={2016},
  publisher={IOP Publishing}
}

@article{chowdhury2020time,
  title={Time-frequency analysis, denoising, compression, segmentation, and classification of PCG signals},
  author={Chowdhury, Tanzil Hoque and Poudel, Khem Narayan and Hu, Yating},
  journal={IEEE Access},
  volume={8},
  pages={160882--160890},
  year={2020},
  publisher={IEEE}
}

@article{ergen2012time,
  title={Time--frequency analysis of phonocardiogram signals using wavelet transform: a comparative study},
  author={Ergen, Burhan and Tatar, Yetkin and Gulcur, Halil Ozcan},
  journal={Computer Methods in Biomechanics and Biomedical Engineering},
  volume={15},
  number={4},
  pages={371--381},
  year={2012},
  publisher={Taylor \& Francis}
}

@article{Dornbush,
  title={Physiology, Heart Sounds},
  author={Sean Dornbush, Andre E. Turnquest},
  journal={National Library of Medicine},
  volume={},
  pages={},
  year={2023},
  publisher={StatPearls Publishing},
doi = {https://www.ncbi.nlm.nih.gov/books/NBK541010/}
}

@article{oliveira2021circor,
  title={The CirCor DigiScope dataset: from murmur detection to murmur classification},
  author={Oliveira, Jorge and Renna, Francesco and Costa, Paulo Dias and Nogueira, Marcelo and Oliveira, Cristina and Ferreira, Carlos and Jorge, Al{\'\i}pio and Mattos, Sandra and Hatem, Thamine and Tavares, Thiago and others},
  journal={IEEE Journal of Biomedical and Health Informatics},
  volume={26},
  number={6},
  pages={2524--2535},
  year={2021},
  publisher={IEEE}
}

@article{atbi2013separation,
  title={Separation of heart sounds and heart murmurs by Hilbert transform envelogram},
  author={Atbi, A and Debbal, SM and Meziani, F and Meziane, A},
  journal={Journal of Medical Engineering \& Technology},
  volume={37},
  number={6},
  pages={375--387},
  year={2013},
  publisher={Taylor \& Francis}
}

@book{cohen2013applied,
  title={Applied Multiple Regression/Correlation Analysis for the Behavioral Sciences},
  author={Cohen, Jacob and Cohen, Patricia and West, Stephen G and Aiken, Leona S},
  year={2013},
  publisher={Routledge}
}

@article{arslan2022automated,
  title={Automated detection of heart valve disorders with time-frequency and deep features on \textsc{PCG} signals},
  author={Arslan, {\"O}zkan},
  journal={Biomedical Signal Processing and Control},
  volume={78},
  pages={103929},
  year={2022},
  publisher={Elsevier}
}

@article{hazeri2021classification,
  title={Classification of normal/abnormal \textsc{PCG} recordings using a time--frequency approach},
  author={Hazeri, Hanie and Zarjam, Pega and Azemi, Ghasem},
  journal={Analog Integrated Circuits and Signal Processing},
  volume={109},
  number={2},
  pages={459--465},
  year={2021},
  publisher={Springer}
}

@article{kui2021heart,
  title={Heart sound classification based on log \textsc{M}el-frequency spectral coefficients features and convolutional neural networks},
  author={Kui, Haoran and Pan, Jiahua and Zong, Rong and Yang, Hongbo and Wang, Weilian},
  journal={Biomedical Signal Processing and Control},
  volume={69},
  pages={102893},
  year={2021},
  publisher={Elsevier}
}

@article{ismail2023pcg,
  title={\textsc{PCG} classification through spectrogram using transfer learning},
  author={Ismail, Shahid and Ismail, Basit and Siddiqi, Imran and Akram, Usman},
  journal={Biomedical Signal Processing and Control},
  volume={79},
  pages={104075},
  year={2023},
  publisher={Elsevier}
}

@article{meziani2012analysis,
  title={Analysis of phonocardiogram signals using wavelet transform},
  author={Meziani, F and Debbal, SM and Atbi, A},
  journal={Journal of Medical Engineering \& Technology},
  volume={36},
  number={6},
  pages={283--302},
  year={2012},
  publisher={Taylor \& Francis}
}

@article{Ergen01042012,
author = {Burhan Ergen and Yetkin Tatar and Halil Ozcan Gulcur},
title = {Time-frequency analysis of phonocardiogram signals using wavelet transform: a comparative study},
journal = {Computer Methods in Biomechanics and Biomedical Engineering},
volume = {15},
number = {4},
pages = {371--381},
year = {2012},
publisher = {Taylor \& Francis},
doi = {10.1080/10255842.2010.538386}
}

@article{benitez2001use,
  title={The use of the Hilbert transform in \textsc{ECG} signal analysis},
  author={Benitez, Diego and Gaydecki, PA and Zaidi, Adnan and Fitzpatrick, AP},
  journal={Computers in Biology and Medicine},
  volume={31},
  number={5},
  pages={399--406},
  year={2001},
  publisher={Elsevier}
}

@article{beyramienanlou2017shannon,
  title={Shannon’s energy based algorithm in \textsc{ECG} signal processing},
  author={Beyramienanlou, Hamed and Lotfivand, Nasser},
  journal={Computational and Mathematical Methods in Medicine},
  volume={2017},
  number={1},
  pages={8081361},
  year={2017},
  publisher={Wiley Online Library}
}

@article{mancia20132013,
  title={2013 ESH/ESC Guidelines for the management of arterial hypertension},
  author={Mancia, Giuseppe and Fagard, Robert and Narkiewicz, Krzysztof and Redon, Josep and Zanchetti, Alberto and B{\"o}hm, Michael and Christiaens, Thierry and Cifkova, Renata and De Backer, Guy and Dominiczak, Anna and others},
  journal={Arterial Hypertension},
  volume={17},
  number={2},
  pages={69--168},
  year={2013}
}

@article{shimamoto2014japanese,
  title={The Japanese Society of Hypertension guidelines for the management of hypertension (JSH 2014)},
  author={Shimamoto, Kazuaki and Ando, Katsuyuki and Fujita, Toshiro and Hasebe, Naoyuki and Higaki, Jitsuo and Horiuchi, Masatsugu and Imai, Yutaka and Imaizumi, Tsutomu and Ishimitsu, Toshihiko and Ito, Masaaki and others},
  journal={Hypertension Research},
  volume={37},
  number={4},
  pages={253--390},
  year={2014},
  publisher={Nature Publishing Group}
}

@article{chadachan2018understanding,
  title={Understanding short-term blood-pressure-variability phenotypes: from concept to clinical practice},
  author={Chadachan, Veerendra Melagireppa and Ye, Min Tun and Tay, Jam Chin and Subramaniam, Kannan and Setia, Sajita},
  journal={International journal of general medicine},
  pages={241--254},
  year={2018},
  publisher={Taylor \& Francis}
}

@misc{AHA_MeasureBP_Accurately,
  author = {{American Heart Association}},
  title  = {Measuring blood pressure accurately},
  note   = {[Online]. Available: \url{https://www2.heart.org/site/DocServer/Break_1_-_Measure_BP_Accurately.pdf}. Accessed: Jan. 8, 2026},
  year   = {2026}
}

@incollection{noble1990electrocardiography,
  author    = {Noble, R. J. and Hillis, J. S. and Rothbaum, D. A.},
  title     = {Electrocardiography},
  booktitle = {Clinical Methods: The History, Physical, and Laboratory Examinations},
  editor    = {Walker, H. K. and Hall, W. D. and Hurst, J. W.},
  edition   = {3},
  chapter   = {33},
  address   = {Boston},
  publisher = {American Medical Association},
  year      = {1990},
  note      = {Chapter 33}
}

@article{surawicz2009aha,
  title={AHA/ACCF/HRS recommendations for the standardization and interpretation of the electrocardiogram: part III: intraventricular conduction disturbances a scientific statement from the American Heart Association Electrocardiography and Arrhythmias Committee, Council on Clinical Cardiology; the American College of Cardiology Foundation; and the Heart Rhythm Society endorsed by the International Society for Computerized Electrocardiology},
  author={Surawicz, Borys and Childers, Rory and Deal, Barbara J and Gettes, Leonard S},
  journal={Journal of the American College of Cardiology},
  volume={53},
  number={11},
  pages={976--981},
  year={2009},
  publisher={American College of Cardiology Foundation Washington, DC}
}

@article{eldakhly2025optimized,
  title={Optimized machine learning for real-time, non-invasive blood pressure monitoring},
  author={Eldakhly, Nabil M},
  journal={The Journal of Supercomputing},
  volume={81},
  number={7},
  pages={1--42},
  year={2025},
  publisher={Springer}
}

@article{springer2015logistic,
  title={Logistic regression-HSMM-based heart sound segmentation},
  author={Springer, David B and Tarassenko, Lionel and Clifford, Gari D},
  journal={IEEE transactions on biomedical engineering},
  volume={63},
  number={4},
  pages={822--832},
  year={2015},
  publisher={IEEE}
}

@inproceedings{dos2025towards,
  title={Towards Non-Intrusive Blood Pressure Estimation Using Thigh ECG and PPG Signals Acquired from a Smart Toilet Seat},
  author={dos Santos Silva, Aline and Correia, Miguel Velhote and da Costa, Andreia Gon{\c{c}}alves and da Silva, Hugo Pl{\'a}cido},
  booktitle={2025 IEEE 8th Portuguese Meeting on Bioengineering (ENBENG)},
  pages={121--124},
  year={2025},
  organization={IEEE}
}

@article{lin2025portable,
  title={Portable ECG and PCG wireless acquisition system and multiscale CNN feature fusion Bi-LSTM network for coronary artery disease diagnosis},
  author={Lin, Junye and Wang, Shaokui and Xuan, Weipeng and Chen, Ding and Liu, Fuhai and Chen, Jinkai and Xia, Shudong and Dong, Shurong and Luo, Jikui},
  journal={Computers in Biology and Medicine},
  volume={191},
  pages={110202},
  year={2025},
  publisher={Elsevier}
}

@article{fynn2025practicality,
  title={Practicality meets precision: Wearable vest with integrated multi-channel PCG sensors for effective coronary artery disease pre-screening},
  author={Fynn, Matthew and Mandana, Kayapanda and Rashid, Javed and Nordholm, Sven and Rong, Yue and Saha, Goutam},
  journal={Computers in Biology and Medicine},
  volume={189},
  pages={109904},
  year={2025},
  publisher={Elsevier}
}

@incollection{chaudhary2025preprocessing,
  title={Preprocessing Techniques for Brain Signal Data: Noise Reduction and Artifact Removal Methods},
  author={Chaudhary, Ujwal},
  booktitle={Expanding Senses using Neurotechnology: Volume 1--Foundation of Brain-Computer Interface Technology},
  pages={119--178},
  year={2025},
  publisher={Springer}
}

@article{angelucci2025wearable,
  title={Wearable devices for patient monitoring in the intensive care unit},
  author={Angelucci, Alessandra and Greco, Massimiliano and Cecconi, Maurizio and Aliverti, Andrea},
  journal={Intensive Care Medicine Experimental},
  volume={13},
  number={1},
  pages={26},
  year={2025},
  publisher={Springer}
}

@article{farrokhi2025reliable,
  title={Reliable peak detection and feature extraction for wireless electrocardiograms},
  author={Farrokhi, Sajad and Dargie, Waltenegus and Poellabauer, Christian},
  journal={Computers in Biology and Medicine},
  volume={185},
  pages={109478},
  year={2025},
  publisher={Elsevier}
}

@article{rabbani2011r,
  title={R peak detection in electrocardiogram signal based on an optimal combination of wavelet transform, hilbert transform, and adaptive thresholding},
  author={Rabbani, Hossein and Mahjoob, M Parsa and Farahabadi, E and Farahabadi, A},
  journal={Journal of Medical Signals \& Sensors},
  volume={1},
  number={2},
  pages={91--98},
  year={2011},
  publisher={Medknow}
}

@article{ergen2013comparison,
  title={Comparison of wavelet types and thresholding methods on wavelet based denoising of heart sounds},
  author={Ergen, Burhan},
  journal={Journal of Signal and Information Processing},
  volume={4},
  number={3B},
  pages={164},
  year={2013},
  publisher={Scientific Research Publishing}
}

@article{liu2018automatic,
  title={An automatic segmentation method for heart sounds},
  author={Liu, Qingshu and Wu, Xiaomei and Ma, Xiaojing},
  journal={Biomedical engineering online},
  volume={17},
  number={1},
  pages={106},
  year={2018},
  publisher={Springer}
}

@inproceedings{strazza2018pcg,
  title={PCG-Delineator: an efficient algorithm for automatic heart sounds detection in fetal phonocardiography},
  author={Strazza, Annachiara and Sbrollini, Agnese and Di Battista, Valeria and Ricci, Rita and Trillini, Letizia and Marcantoni, Ilaria and Morettini, Micaela and Fioretti, Sandro and Burattini, Laura},
  booktitle={2018 Computing in Cardiology Conference (CinC)},
  volume={45},
  pages={1--4},
  year={2018},
  organization={IEEE}
}

@book{pachori2023time,
  title={Time-frequency analysis techniques and their applications},
  author={Pachori, Ram Bilas},
  year={2023},
  publisher={CRC Press}
}

@article{giordano2021automated,
  title={Automated assessment of the quality of phonocardographic recordings through signal-to-noise ratio for home monitoring applications},
  author={Giordano, Noemi and Rosati, Samanta and Knaflitz, Marco},
  journal={Sensors},
  volume={21},
  number={21},
  pages={7246},
  year={2021},
  publisher={MDPI}
}

@article{damani2021investigation,
  title={Investigation of synchronized acquisition of electrocardiogram and phonocardiogram signals towards electromechanical profiling of the heart},
  author={Damani, Devanshi N and Sundaram, DivaaNar Siva Baala and Damani, Shivam and Kapoor, AnoushNa and Arruda-Olson, Adelaide M and Arunachalam, Shivaram P},
  journal={Biomed Sci Instrum},
  volume={57},
  pages={2},
  year={2021}
}

@article{loro2024validation,
  title={Validation of a wearable sensor prototype for measuring heart rate to prescribe physical activity: cross-sectional exploratory study},
  author={Loro, Fernanda La{\'\i}s and Martins, Riane and Ferreira, Jana{\'\i}na Barcellos and de Araujo, Cintia Laura Pereira and Prade, Lucio Rene and Both, Cristiano Bonato and Nobre, J{\'e}ferson Campos Nobre and Monteiro, Mariane Borba and Dal Lago, Pedro},
  journal={JMIR Biomedical Engineering},
  volume={9},
  pages={e57373},
  year={2024},
  publisher={JMIR Publications Toronto, Canada}
}

@article{adams2022accurate,
  title={Accurate detection of heart rate using in-ear photoplethysmography in a clinical setting},
  author={Adams, Tim and Wagner, Sophie and Baldinger, Melanie and Zellhuber, Incinur and Weber, Michael and Nass, Daniel and Surges, Rainer},
  journal={Frontiers in Digital Health},
  volume={4},
  pages={909519},
  year={2022},
  publisher={Frontiers Media SA}
}

@inproceedings{prasad2020detection,
  title={Detection of S1 and S2 locations in phonocardiogram signals using zero frequency filter},
  author={Prasad, RaviShankar and Yilmaz, Gurkan and Chetelat, Olivier and Doss, Mathew Magimai-},
  booktitle={ICASSP 2020-2020 IEEE International Conference on Acoustics, Speech and Signal Processing (ICASSP)},
  pages={1254--1258},
  year={2020},
  organization={IEEE}
}

@inproceedings{yamacli2008segmentation,
  title={Segmentation of S1--S2 sounds in phonocardiogram records using wavelet energies},
  author={Yamacli, Mustafa and Dokur, Zumray and Olmez, Tamer},
  booktitle={2008 23rd International Symposium on Computer and Information Sciences},
  pages={1--6},
  year={2008},
  organization={IEEE}
}

@article{arjoune2024noise,
  title={A noise-robust heart sound segmentation algorithm based on Shannon energy},
  author={Arjoune, Youness and Nguyen, Trong N and Doroshow, Robin W and Shekhar, Raj},
  journal={IEEE Access},
  volume={12},
  pages={7747--7761},
  year={2024},
  publisher={IEEE}
}

@article{coste2025comparative,
  title={A comparative study between \textsc{ECG}-and \textsc{PPG}-based heart rate sensors for heart rate variability measurements: influence of body position, duration, sex, and age},
  author={Coste, Alexandre and Millour, Geoffrey and Hausswirth, Christophe},
  journal={Sensors},
  volume={25},
  number={18},
  pages={5745},
  year={2025},
  publisher={MDPI}
}

@article{ganti2020wearable,
  title={Wearable cuff-less blood pressure estimation at home via pulse transit time},
  author={Ganti, Venu G and Carek, Andrew M and Nevius, Brandi N and Heller, J Alex and Etemadi, Mozziyar and Inan, Omer T},
  journal={IEEE journal of biomedical and health informatics},
  volume={25},
  number={6},
  pages={1926--1937},
  year={2020},
  publisher={IEEE}
}

@article{omari2019new,
  title={A new approach for blood pressure estimation based on phonocardiogram},
  author={Omari, Tahar and Bereksi-Reguig, Fethi},
  journal={Biomedical engineering letters},
  volume={9},
  number={3},
  pages={395--406},
  year={2019},
  publisher={Springer}
}

@inproceedings{dia2019fetal,
  title={Fetal heart rate estimation from a single phonocardiogram signal using non-negative matrix factorization},
  author={Dia, Nafissa and Fontecave-Jallon, Julie and Gumery, Pierre-Yves and Rivet, Bertrand},
  booktitle={2019 41st Annual International Conference of the IEEE Engineering in Medicine and Biology Society (EMBC)},
  pages={5983--5986},
  year={2019},
  organization={IEEE}
}

%
\end{document}